\documentclass[12pt,preprint]{aastex}

\def\be{\begin{equation}}
\def\ee{\end{equation}}

\def\be{\begin{equation}}
\def\ee{\end{equation}}
\catcode`\@=11 
\def\@versim#1#2{\vcenter{\offinterlineskip
\ialign{$\m@th#1\hfil##\hfil$\crcr#2\crcr\sim\crcr } }}
\def\lsim{\mathrel{\mathpalette\@versim<}}
\def\gsim{\mathrel{\mathpalette\@versim>}}

\newcommand{\mev}{ {\rm MeV} }

\newcommand{\g}{ {\rm g} }
\newcommand{\cm}{ {\rm cm} }
\newcommand{\s}{ {\rm s} }
\newcommand{\K}{ {\rm K} }

\newcommand{\erg}{ {\rm erg} }
\newcommand{\mbh}{ {M_{\rm BH}} }

\newcommand{\anti}[1]{{\overline{#1}}}
\newcommand{\gtsima}{$\; \buildrel > \over \sim \;$}
\newcommand{\ltsima}{$\; \buildrel < \over \sim \;$}
\newcommand{\simgt}{\lower.5ex\hbox{\gtsima}}
\newcommand{\simlt}{\lower.5ex\hbox{\ltsima}}

\usepackage{epsf,graphics}
\shorttitle{NDAF and supernovae}
\shortauthors{Kohri et al.}
\begin{document}

\title{Neutrino-Dominated Accretion and Supernovae}

\author{Kazunori Kohri$^{\dagger, \star}$}
\author{Ramesh Narayan$^{\dagger}$}
\author{Tsvi Piran$^{**,\ddag}$}
\affil{
$^\dagger$Harvard-Smithsonian Center for Astrophysics, 60 Garden Street,
Cambridge, MA 02138}

\affil{
$^\star$Department of Earth and Space Science,
Osaka University, Toyonaka 560-0043,
Japan
}

\affil{
$^{**}$Racah Institute for Physics, Hebrew University, Jerusalem 91904, Israel
}
\affil{
$^{\ddag}$Theoretical Astrophysics Caltech, Pasadena, CA 91125,
USA }
\email{kkohri@cfa.harvard.edu}

\begin{abstract}
We suggest that part of the infalling material during the
core-collapse of a massive star goes into orbit around the compact
core to form a hot, dense, centrifugally-supported accretion disk
whose evolution is strongly influenced by neutrino interactions.
Under a wide range of conditions, this neutrino-dominated accretion
flow will be advection-dominated and will develop a substantial
outflowing wind. We estimate the energy carried out in the wind and
find that it exceeds $10^{50}$ erg for a wide range of parameters and
even exceeds $10^{51}$ erg for reasonable parameter choices.  We
propose that the wind energy will revive a stalled shock and will help
produce a successful supernova explosion.  We discuss the role of the
disk wind in both prompt and delayed explosions. While both scenarios
are feasible, we suggest that a delayed explosion is more likely, and
perhaps even unavoidable.  Finally, we suggest that the disk wind may
be a natural site for $r$-process nucleosynthesis.

\end{abstract}


\section{Introduction}

Almost forty years after the original computations of Colgate \& White
(1966) and Arnett (1967), it is still not known exactly how core
collapse supernovae explode. Quite early on it became clear that the
prompt shock that follows the bounce of the core stagnates, leading to
a standing shock at a distance of several hundred km from the
center. The region interior to the shock contains hot material that
settles onto the central core, while the exterior contains infalling
matter.  Unless some additional energy source is available to
energize the post-shock gas, the shock will ultimately run out of
energy and we would have a failed explosion.

During the mid eighties, Wilson (1985) and Bethe \& Wilson (1985; see
also Goodman, Dar \& Nussinov 1987) suggested that late time neutrino
heating, about a second after the bounce, could lead to a revival of
the shock and to ejection of the infalling envelope. However, later
calculations indicated that this late neutrino heating might be
insufficient. The fact that the explosion energy (a few foe $\equiv 10^{51}$
erg) is much smaller than the binding energy of the neutron star (a
few hundred foe) gave some hope that small details in the calculations
may change the result, leaving open the possibility that yet more
sophisticated calculations might ultimately produce an
explosion. However, recent detailed computations, carried out by
different groups with improved neutrino transport and fully
relativistic hydrodynamics (Bruenn 1985; Bruenn \& Mezzacappa 1997;
Bruenn, De Nisco, \& Mezzacappa 2001; Liebendorfer et al. 2001, 2004;
Mezzacappa, et al. 2001; Myra \& Burrows, 1990; Rampp \& Janka 2000;
Thompson, Burrows \& Pinto 2003), have demonstrated conclusively that
one-dimensional, i.e., spherically-symmetric, simulations do not lead
to a supernova (see Burrows \& Thompson, 2002 for a recent review). A
consensus has emerged that the coupling efficiency of the emerging
neutrinos to the mantle is too small to lead to an explosion, so that
the failure to explode is not an artifact of inaccurate calculations.
The neutrinos simply do not transfer enough energy to the mantle, and
something else is required beyond one-dimensional collapse and
neutrino transport (see e.g.  Burrows \& Thompson, 2002).

Two-dimensional calculations carried out in the mid nineties
(Herant et al. 1994; Burrows, Hayes \& Fryxell, 1995; Janka \&
Muller, 1996; Fryer et al. 1999) resulted in successful
supernovae.  These calculations also revealed turbulence in the
collapsing gas, and it was suggested that the explosion was the
result of an enhancement of the coupling between the neutrinos and
the mantle due to the turbulence (Burrows \& Thompson, 2002).
However, it is unclear at present whether or not the results are
an artifact of the simplified transport models used in the
simulations.  It is also unclear if the two-dimensional turbulence
seen in the calculations provides a realistic description of full
three-dimensional turbulence (see e.g., Fryer \& Warren 2002,
2004).

In view of this confusing situation, it would be wise to search
for additional energy sources that might transport energy to the
mantle and lead to an explosion. Burrows \& Goshy (1993) describe
the explosion problem in terms of the neutrino luminosity needed
to produce an explosion at a given mass accretion rate. Clearly,
what is important is the energy flux, whether it is via neutrinos
or something else.  In fact, energy in a form other than neutrinos
is likely to be superior since it might couple more efficiently to
the matter in the mantle\footnote{The efficiency of capturing
neutrinos is usually considered to be less than a percent.  But
this is the ratio of the neutrinos captured to the total number of
neutrinos radiated. However, the neutrino emission lasts for ten
seconds while the delayed shock phase lasts only one second. What
is more relevant for our comparison is the capture efficiency of a
few percent that corresponds to the ratio of neutrinos captured to
the neutrinos radiated during the delayed shock phase --- a few
$\times 10^{52}$ erg (Burrows \& Thompson 2002). }.  

Two energy sources that have been mentioned a number of times in
connection with the supernova problem are magnetic fields (Le Blanc \&
Wilson, 1970; Moiseenko, Bisnovatyi-Kogan \& Ardeljan 2004; Kotake et
al.  2004; Wheeler, Meier \& Wilson 2002) and rotation.  Rotation has
been discussed as the cause of asymmetry in the explosion and also in
the context of generating gravitational waves.  Recent progress in
numerical computations have enabled modeling rotating collapse with a
realistic equation of state and neutrino transport in two dimensions
(Fryer \& Heger 2000; Kotake, Yamada \& Sato 2003; Livne et al.,2004;
Walder et al. 2004; and references therein) and even three dimensions
(Fryer \& Warren 2004; Janka et al. 2004).  Some of these studies
indicate that rotation may cause a weakening of two-dimensional
convection and thereby delay the explosion (Fryer \& Heger, 2000).
However, more work is needed before one can be sure of the net effect
of rotation on supernova explosions.

We discuss in this paper a novel scenario, based on rotation, which
does not seem to have been considered before. We suggest that a
core-collapse supernova may be partially driven by wind energy from an
accretion disk that forms around the proto-neutron star (inside the
stalled shock). Such a disk is expected to form if the pre-explosion
stellar core rotates reasonably rapidly.  The required progenitor
rotation rate for our scenario to work is comparable to, or perhaps
even larger than, the rates predicted in current stellar evolution
models (e.g., Heger et al. 2000, 2004).  However, there are large
uncertainties in the stellar model calculations.

In our model, because of the high angular momentum of the stellar gas,
the infalling matter arranges itself into a dense and very hot disk in
the form of a Neutrino Dominated Accretion Flow (NDAF; Popham, Woosley
\& Fryer 1999; Narayan, Piran \& Kumar 2001; Kohri \& Mineshige 2002;
Di Matteo, Perna \& Narayan 2002)~\footnote{For a numerical simulation
of  the dynamical evolution of an NDAF, see Lee, Ramirez-Ruiz \& Page
(2004).}. For a wide range of parameter space, an NDAF is
advection-dominated and is expected to emit a substantial wind. We
propose that the mechanical energy transported out by this wind will
be transferred to the gas inside the stalled shock, reviving the shock
and leading to a successful explosion. Related ideas have been
discussed by MacFadyen (2003) and Thompson, Quataert \& Burrows
(2004). Note that the wind that we consider originates from the disk
and is different from the spherical wind originating from the
proto-neutron star itself discussed by Thompson et al. (2000).

\S~2 of the paper presents a detailed discussion of the structure of
the NDAF that forms during core collapse. The model we employ for the
calculations is more elaborate than those described in the recent
literature (Popham et al. 1999; Narayan et al. 2001; Kohri \&
Mineshige 2002; Di Matteo et al. 2002), and we highlight the
improvements. Some of the relevant physics is described in
Appendices~A and B.  These structure equations are applicable to an
NDAF residing in the core of a collapsar (MacFadyen \& Woosley 1999)
and may be of use for numerical simulations of collapsars. The reader
who is not interested in the technical details and wishes to read
about the application to supernovae is invited to skip \S~2 and to go
directly to \S~3, where we calculate the mechanical energy that is
likely to be carried in the disk wind. Even with fairly conservative
assumptions, we show that up to about $10^{51}$ erg of energy might be
available in the wind.  We suggest that this energy would help to
energize the stalled post-bounce shock in a supernova. The material in
the wind may also participate in $r$-process nucleosynthesis.  \S~4
concludes with a discussion.

\section{Model of a Neutrino Dominated Accretion Disk}

\subsection{Physical variables}

In this paper we use the following dimensionless variables for the
mass of the compact star $M$, the accretion rate $\dot{M}$ and the
radius $R$:
\begin{eqnarray}
    \label{eq:smallm}
    m \equiv M/M_{\odot},
\end{eqnarray}
\begin{eqnarray}
    \label{eq:smallmdot}
    \dot{m} \equiv \dot{M}/(M_{\odot}~\s^{-1}),
\end{eqnarray}
\begin{eqnarray}
    \label{eq:smallr}
    r \equiv R/R_{\rm S} = R/(2.95\times10^{5} m \ \cm),
\end{eqnarray}
where $R_{\rm S}$ is the Schwarzschild radius.

We also scale the matter density $\rho$ and the temperature $T$ as
follows:
\begin{eqnarray}
    \label{eq:rho10}
    \rho_{10} \equiv \rho/(10^{10} \g \ \cm^{-3}),
\end{eqnarray}
\begin{eqnarray}
    \label{eq:T11}
    T_{11} \equiv T/(10^{11} \K).
\end{eqnarray}
We write the surface density of the disk as
\begin{eqnarray}
    \label{eq:sigma}
    \Sigma = 2 H \rho,
\end{eqnarray}
where the disk half-thickness (or disk height) $H$ is given by
\begin{eqnarray}
    \label{eq:Height}
    H = c_{S}/\Omega_{K}.
\end{eqnarray}
Here $c_{S}$ is the sound speed, which is defined by
\begin{eqnarray}
    \label{eq:sound-speed}
    c_{S} = \sqrt{p/\rho},
\end{eqnarray}
and $\Omega_K$ the Keplerian angular velocity, given by
\begin{eqnarray}
    \label{eq:Omega_K}
    \Omega_{K} = \sqrt{G M/R^{3}} = 17.19 \times 10^{4} m^{-1} r^{3/2}~\s^{-1}.
\end{eqnarray}
The pressure $p$ is the sum of contributions from radiation,
electrons/positrons, nuclei, and neutrinos:
\begin{eqnarray}
    \label{eq:p_total}
    p = p_{\rm rad} + p_{e} + p_{\rm gas} + p_{\nu}
\end{eqnarray}
In the next subsection, we discuss each component of the pressure in
detail.

\subsection{Contributions to the Pressure}
\label{subsec:press}

The radiation pressure is given by
\begin{eqnarray}
    \label{eq:p_rad}
    p_{\rm rad} = \frac13 a T^{4} = 2.52 \times 10^{29}~\erg~\cm^{-3}
    T_{11}^{4},
\end{eqnarray}
where $a$ is the radiation density constant.  The gas pressure (from
nuclei) is
\begin{eqnarray}
    \label{eq:p_gas}
    p_{\rm gas} = \rho k_{\rm B} T/m_{N} = 3.26 \times 10^{28}~\erg~\cm^{-3}
    \rho_{10} T_{11},
\end{eqnarray}
where $m_{N}$ is the mean mass of nuclei.  We are mostly interested in
fully dissociated nuclei, so $m_{N}$ is the mass of a nucleon, where
we ignore the mass difference between a proton and a neutron.

\begin{figure}
\epsscale{1.0}
\plotone{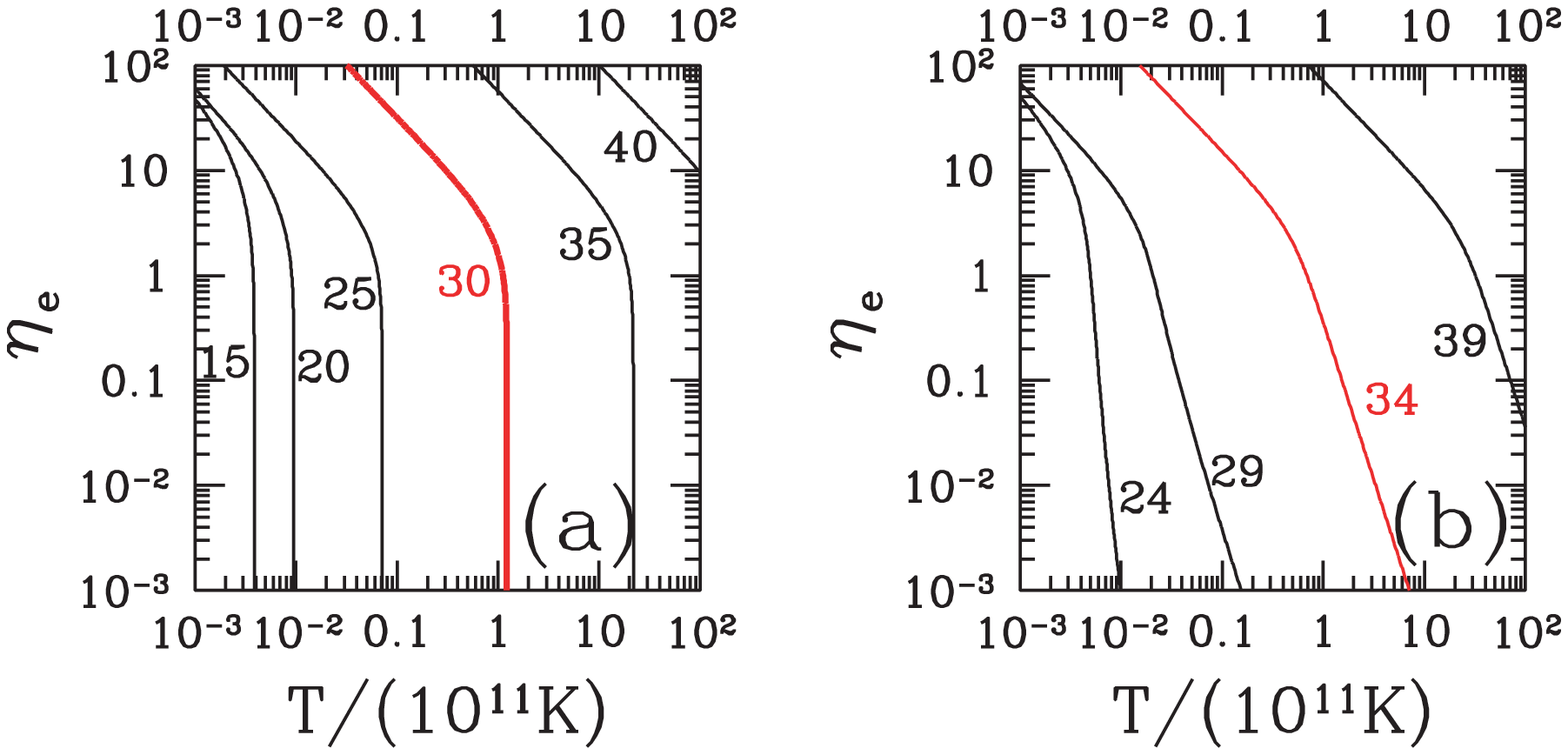}
\vspace{.0in}
\caption{(a) Contours of the total pressure of electrons and
positrons, $p_{e}\equiv p_{e^-} + p_{e^+}$
(eqs. \ref{eq:epressure}--\ref{eq:posipress}), in the $T$--$\eta_{e}$
plane. The contours are labeled by the value of
$\log_{10}[p_{e}/(\erg~\cm^{-3})]$.  (b) Contours of the net number
density of electrons, $n_{e}\equiv n_{e^-} - n_{e^+}$
(eqs. \ref{eq:nete}--\ref{eq:n_+}), in the $T$--$\eta_e$ plane.  The
contours are labeled by the value of $\log_{10}[n_{e}/(\cm^{-3})]$.}
\label{fig:penepanel}
\end{figure}

For the electron pressure, it is important to consider a general form
that is valid even when electrons are degenerate and/or
relativistic. We write
\begin{eqnarray}
    \label{eq:epressure}
    p_{e} = p_{e^{-}} + p_{e^{+}},
\end{eqnarray}
where the two terms represent the contributions from electrons and
positrons, given by
\begin{eqnarray}
    \label{eq:elepress}
    p_{e^{-}}= \frac{1}{3\pi^{2}\hbar^{3}c^{3}}\int^{{\infty}}_{0}
    dp \frac{p^{4}}{\sqrt{p^{2}c^{2}+m_{e}^{2}c^{4}}}
    \frac{1}{e^{(\sqrt{p^{2}c^{2}+m_{e}^{2}c^{4}}-\mu_{e})/k_{\rm
    B}T} + 1},
\end{eqnarray}
\begin{eqnarray}
    \label{eq:posipress}
    p_{e^{+}}= \frac{1}{3\pi^{2}\hbar^{3}c^{3}}\int^{{\infty}}_{0}
    dp \frac{p^{4}}{\sqrt{p^{2}c^{2}+m_{e}^{2}c^{4}}}
    \frac{1}{e^{(\sqrt{p^{2}c^{2}+m_{e}^{2}c^{4}}+\mu_{e})/k_{\rm
    B}T} + 1},
\end{eqnarray}
$\mu_e$ is the chemical potential of electrons, $c$ is the speed
of light, and $m_e$ is the electron mass. Note that the above
expression is applicable for both relativistic and nonrelativistic
electrons.  The chemical potential $\mu_{e}$ has to be determined
self-consistently, as we discuss in detail in the next two
subsections. In Fig.~\ref{fig:penepanel}~(a), we show contours of
$p_{e}$ in units of $\erg~\cm^{-3}$ in the $T$--$\eta_{e}$ plane,
where $\eta_{e}$ is the degeneracy parameter of electrons, defined by
\begin{eqnarray}
    \label{eq:eta_e}
    \eta_e = \mu_e / (k_{\rm B}T).
\end{eqnarray}
This parameter is a good measure of the electron degeneracy.  If
$\eta_{e}$ is much larger than unity, the electrons are strongly
degenerate, whereas, if $\eta_{e} \ll 1$, the electrons are weakly
degenerate and we can ignore degeneracy.

The last term in Eq.~(\ref{eq:p_total}) is the pressure of the
neutrinos, $p_{\nu}=\sum_{i=e,\mu,\tau} p_{\nu_{i}}$, for the three
species $\nu_{i}=$ $\nu_{e}$, $\nu_{\mu}$, $\nu_{\tau}$. Adopting the
approximate formula in Popham \& Narayan (1995) and Di Matteo, Perna
\& Narayan (2002), we write
\begin{eqnarray}
    \label{eq:p_nu}
    p_{\nu_{i}} &=& \frac13 u_{\nu_{i}} = \frac13 \ \frac{( 7/8) a T^{4}
    \left( \tau_{\nu_{i}}/2 +
    1/\sqrt{3} \right)}{\tau_{\nu_{i}}/2+1/\sqrt{3}+1/ (3\tau_{a,
    \nu_{i}})}  \nonumber \\
    &=&
    2.21\times10^{29}~\erg~\cm^{-3}T_{11}^{4}\ \frac{\tau_{\nu_{i}}/2 +
    1/\sqrt{3}}{\tau_{\nu_{i}}/2+1/\sqrt{3}+1/ (3\tau_{a,
    \nu_{i}})}.
\end{eqnarray}
The various neutrino ``optical depths'' $\tau$ are discussed in detail
in subsection~\ref{subsec:nucool} (see,
Eqs.~(\ref{eq:tauae}),~(\ref{eq:tauamu})~and~(\ref{eq:tause_nuN})).

\subsubsection{Chemical potential of electrons}

When electrons are degenerate, some important effects are
introduced. First, the process of neutrino emission is considerably
modified; for example, pair creation of neutrinos and antineutrinos is
suppressed because of the asymmetry between electrons and
positrons. In addition, as we have discussed above, the expression for
the electron pressure is modified.

The asymmetry between electrons and positrons is characterized by the
electron chemical potential $\mu_e$, which is determined by the
condition of charge neutrality among protons, electrons and
positrons. Let us introduce the net number density of electrons
$n_{e}$.  Charge neutrality requires
\begin{eqnarray}
    \label{eq:nete}
    n_{e} \equiv n_{e^-} - n_{e^+} = n_p,
\end{eqnarray}
where $n_p$ is the number density of protons.  The number densities of
electrons and positrons are given by
\begin{eqnarray}
    \label{eq:n_-}
    n_{e^-} &=& \frac{1}{\hbar^3\pi^2}\int^{\infty}_{0}dp p^2
    \frac{1}{e^{(\sqrt{p^2c^2 + m_e^2c^4} - \mu_e)/k_{\rm B}T} + 1}, \\
    \label{eq:n_+}
    n_{e^+} &=& \frac{1}{\hbar^3\pi^2}\int^{\infty}_{0}dp p^2
    \frac{1}{e^{(\sqrt{p^2c^2 + m_e^2c^4} + \mu_e)/k_{\rm B}T} + 1}.
\end{eqnarray}
Thus, for given $n_p$ and $T$, we can solve equation (\ref{eq:nete})
to obtain $\mu_e$, and thereby calculate $n_{e^-}$ and $n_{e^+}$.  We
plot contours of $n_{e}$ in the $T$--$\eta_{e}$ plane in
Fig.~\ref{fig:penepanel}~(b).

\subsubsection{Neutron to proton ratio}
\label{subsubsec:n-p-ratio}

To calculate the chemical potential $\mu_e$ for a given matter density
$\rho$ and temperature $T$, we should know the correct value of {\it
the neutron to proton ratio}, $n/p \equiv n_{n}/n_{p}$, since this
determines the value of $n_p$ needed in equation (\ref{eq:nete}). Here
$n_{N}$ ($N=n,p$) is the number density of nucleon $N$. We need $n/p$
also to calculate the neutrino emission rates in
\S~\ref{subsec:nucool}. Therefore, it is quite crucial to obtain $n/p$
as accurately as possible. We briefly discuss here the procedure we
employ, leaving the details to Appendix~\ref{sec:actural_np}.

\begin{figure}
\epsscale{1.0}
\plotone{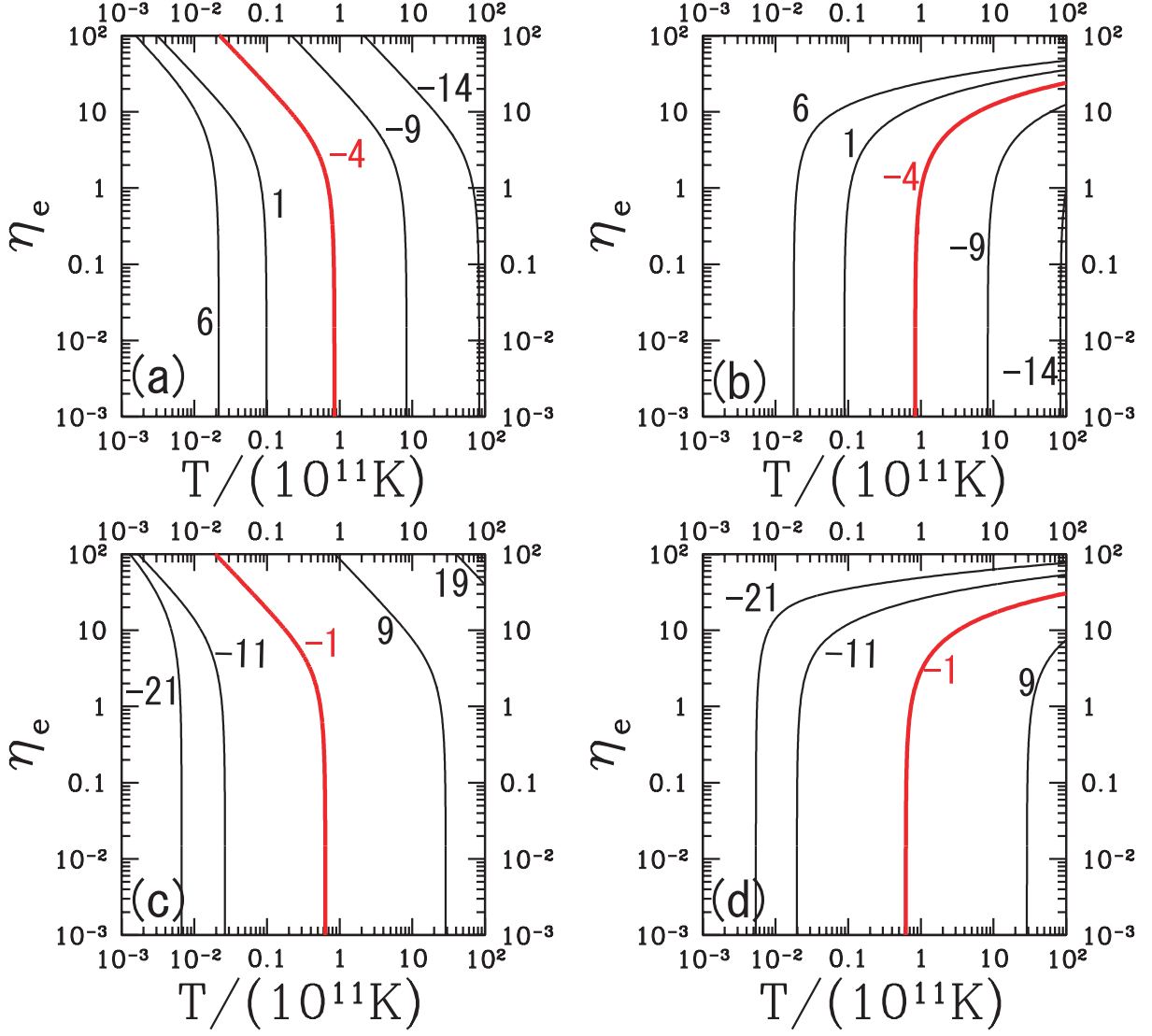}
\vspace{.0in}
\caption{(a) Contours of the proton to neutron conversion timescale
$\log_{10}[\Gamma_{pe^- \to n \nu_e}^{-1}/\s]$
(eq.~\ref{eq:beta_reac2}).  (b) Same as (a), but for the neutron to
proton reaction timescale $\Gamma_{ne^+ \to p \anti{\nu}_e}^{-1}$
(eq.~\ref{eq:beta_reac1}).  (c) Contours of the neutrino-cooling rate
$\log_{10}[q^-_{pe^- \to n\nu_e}/n_p/(\erg~\s^{-1})]$
(eq.~\ref{eq:e-p}) in the $T$--$\eta_{e}$ plane.  (d) Same as (c), but
for the rate $\log_{10}[q^-_{ne^+\to p
\anti{\nu}_e}/n_n/(\erg~\s^{-1})]$ (eq.~\ref{eq:e+n}).  }
\label{fig:amma_q_panel}
\end{figure}

To obtain $n/p$, we need to solve a set of reaction equations between
neutrons and protons.  We consider only the weak interaction; since
the mean energy of nucleons in the system is much lower than the rest
mass energy of pions $m_{\pi}c^{2} \sim$ 140 MeV, we may ignore the
strong interaction.  We assume that the photons and electrons
(including positrons) are completely thermalized since the timescale
of electromagnetic interactions is much shorter than the dynamical
timescale (see Appendix~\ref{sec:timescale}). The reaction equations
between protons and neutrons are then given by
\begin{eqnarray}
    \label{eq:dnpdt}
    \frac{dn_{p}}{dt} = - \Gamma_{p \to n} n_{p} + \Gamma_{n \to p} n_{n},
\end{eqnarray}
\begin{eqnarray}
    \label{eq:dnndt}
    \frac{dn_{n}}{dt} = + \Gamma_{p \to n} n_{p} - \Gamma_{n \to p} n_{n},
\end{eqnarray}
where $\Gamma_{p \to n}$ ($\Gamma_{n \to p}$) is the transition rate
from proton to neutron (neutron to proton). The rate is represented by
the sum of several weak interaction rates: $\Gamma_{p \rightarrow n} =
\Gamma_{pe^- \to n \nu_e}+ \Gamma_{p\anti{\nu}_e \to ne^+} +
\Gamma_{pe^-\anti{\nu}_e \to n}$ ($\Gamma_{n \rightarrow p} =
\Gamma_{ne^+ \to p \anti{\nu}_e}+\Gamma_{n \nu_e \to p e^-} +
\Gamma_{n \to p e^- \anti{\nu}_e}$), with
\begin{eqnarray}
    \label{eq:beta_reac1}
    \Gamma_{ne^+ \to p \anti{\nu}_e}\hspace{-0.3cm}&=&
    \hspace{-0.3cm} K_1  \int_{Q/c+m_ec}^{\infty} \hspace{-0.6cm}
    dp\left[\sqrt{(pc-Q)^2-m_e^2c^{4}}
    (pc-Q)
      \frac{p^2}{e^{(pc - Q+\mu_{e})/k_{\rm B}T}+1}
    \left( 1-f_{\anti{\nu}_e}(p) \right) \right], \\
    \label{eq:beta_reac2}
    \Gamma_{pe^- \to n \nu_e}\hspace{-0.3cm}&=&
    \hspace{-0.3cm} K_1  \int_{0}^{\infty}\hspace{-0.5cm}
    dp\left[\sqrt{(pc+Q)^2-m_e^2c^{4}}
    (Q+pc)
      \frac{p^2}{e^{(pc + Q-\mu_{e})/k_{\rm B}T}+1}
    \left( 1-f_{\nu_e}(p) \right)\right], \\
    \label{eq:beta_reac3}
     \Gamma_{ n \to p e^- \anti{\nu}_e}\hspace{-0.3cm}&=&
     \hspace{-0.3cm} K_1  \int_0^{Q/c-m_ec} \hspace{-0.7cm}
     dp\left[\sqrt{(pc-Q)^2-m_e^2c^{4}}
    (Q-pc)
       \frac{p^2}{1+e^{(pc-Q+\mu_{e})/k_{\rm B}T}}
    \left(1-f_{\anti{\nu}_e}(p)\right) \right], \\
    \label{eq:beta_reac4}
    \Gamma_{n \nu_e \to p e^-}\hspace{-0.3cm}&=&
    \hspace{-0.3cm} K_1  \int_0^{\infty} dp\left[\sqrt{(pc+Q)^2-m_e^2c^{4}}
    (pc+Q)  \frac{p^2}{1+e^{-(pc+Q-\mu_{e})/k_{\rm B}T}}
    f_{\nu_e}(p)\right], \\
    \label{eq:beta_reac5}
    \Gamma_{pe^-\anti{\nu}_e \to n}\hspace{-0.3cm}&=&
    \hspace{-0.3cm} K_1  \int_{0}^{Q/c-m_ec} \hspace{-0.6cm}
    dp\left[\sqrt{(pc-Q)^2-m_e^2c^{4}}
    (Q-pc)
      \frac{p^2}{e^{-(pc - Q+\mu_{e})/k_{\rm B}T}+1}
    f_{\anti{\nu}_e}(p) \right], \\
    \label{eq:beta_reac6}
    \Gamma_{p\anti{\nu}_e \to ne^+}\hspace{-0.3cm}&=&
    \hspace{-0.3cm} K_1
    \int_{Q/c+m_ec}^{\infty}dp\left[\sqrt{(pc-Q)^2-m_e^2c^{4}}
    (Q-pc)
      \frac{p^2}{1+e^{-(pc - Q+\mu_{e})/k_{\rm B}T}}
    f_{\anti{\nu}_e}(p) \right],
\end{eqnarray}
where $f_{\nu_e}$ ($f_{\anti{\nu}_e}$) is the distribution function of
electron neutrinos (antineutrinos), and $Q = (m_n - m_p)c^{2} = 1.29$
MeV. The normalization factor $K_1$ is obtained from the neutron
lifetime $\tau_n $ as $ K_1 =G_{\rm
F}^{2}(1+3g_{A})c/(2\pi^{3}\hbar^{3}) \simeq (1.636 \tau_n)^{-1}$,
where $G_{\rm F}$ is the Fermi coupling constant and $g_A$ is the
axial vector coupling constant.  The present best estimate of the
neutron lifetime is $\tau_n \simeq 885.7 \pm 0.8$, according to a
recent compilation of the experimental data (Eidelman et al. 2004).
For reference, we plot the timescale of $\Gamma_{pe^- \to n
\nu_e}^{-1}$ and $\Gamma_{ne^+ \to p \anti{\nu}_e}^{-1}$ in the
$T$--$\eta_{e}$ plane in Fig.~\ref{fig:amma_q_panel}~(a) and
Fig.~\ref{fig:amma_q_panel}~(b), respectively. These are the two most
important rates for the calculation of $n/p$ (see the discussion in
Appendix~\ref{sec:actural_np}).

The equilibrium value of the neutron to proton ratio is calculated by
imposing the condition $dn_{p}/dt = dn_{n}/dt = 0$. From
Eqs.~(\ref{eq:dnpdt})~and~(\ref{eq:dnndt}), this gives
\begin{eqnarray}
    \label{eq:npratio}
    \left(\frac{n}{p}\right) = \frac{\Gamma_{p \to
    n}}{\Gamma_{{n \to p}}}.
\end{eqnarray}
Note that the electron chemical potential $\mu_{e}$ is implicitly
present in the above relation since the interconverting reaction rates
all depend on it.  On the other hand, the solution for $\mu_e$ via the
charge neutrality condition Eq.~(\ref{eq:nete}) requires the value of
$n/p$.  Thus we have a highly nonlinear coupled system of equations
which has to be solved recursively, given the matter density $\rho$
and the temperature $T$.

To calculate $n/p$ via Eq.~(\ref{eq:npratio}), we need to know the
distribution function of neutrinos $f_{\nu_e}$ and antineutrinos
$f_{\anti{\nu}_e}$.  If the neutrinos are perfectly thermalized and
have an ideal Fermi-Dirac distribution, then from
Eq.~(\ref{eq:npratio}) with
Eqs.~(\ref{eq:beta_reac1})~--~(\ref{eq:beta_reac6}) we obtain a simple
analytic expression for the neutron to proton ratio,
\begin{eqnarray}
    \label{eq:npratio_EQ}
    \left(\frac{n}{p}\right)_{\rm Eq} = \exp\left(- \frac{Q}{k_{\rm B}T} +
    \eta_{e} \right).
\end{eqnarray}
But this result is valid only for perfect thermodynamic
equilibrium~\footnote{For simplicity, we have ignored the chemical
potential of neutrinos. This approximation is reasonable for our
problem because the region where the neutrinosphere appears is
immediately replaced by fresh plasma from the outer optically thin
region (but nucleons are still in beta equilibrium) until a large
amount of lepton asymmetry is produced in the disk. In the optically
thick region, the timescale for order unity change of the electron
neutrino chemical potential by neutrino emission is typically $\sim 1$
sec.  In comparison, the viscous timescale $R/v_r$ and the dynamical
timescale $1/\Omega$ of the accreting matter are less than 0.1 sec
(see \S~3). Note that this is different from the normal case of
spherical collapse in a supernova.  Incidentally, note also that
Eq.~(\ref{eq:npratio_EQ}) is not the observed neutrino spectrum from
the SN; it is the local neutrino distribution. It is well-known that
the observed spectrum is modified from the perfect Fermi-Dirac
distribution by the energy dependence of the cross section of
neutrino-nucleon reactions. Usually, this is fitted by using the
effective chemical potential of the neutrino.}.

In the more general case, when the neutrinos are not fully thermalized,
we have to calculate the correct form of the distribution function of
neutrinos.  The exact way to do this is to solve a set of Boltzmann
equations for the time-evolution and energy transfer. This is a major
exercise which is not necessary for our present purpose.  We have
instead adopted an approximate procedure, which is discussed in
Appendix~\ref{sec:actural_np}.

\subsection{Heating and cooling rates}
\label{sec:heating-cooling-rates}

\subsubsection{Heating rate}

In the theory of accretion disks, the energy balance between heating
and cooling processes plays an important role. The standard thin
accretion disk solution corresponds to the case in which energy loss
by cooling dominates over energy advection, whereas an
advection-dominated accretion flow corresponds to the opposite limit
(Narayan \& Yi 1994, 1995a).

Making use of the standard disk equations (e.g., Frank, King \& Raine
1992), the vertically integrated viscous heating rate (per unit area)
over a half thickness $H$ is given by
\begin{eqnarray}
    \label{eq:Qvis}
    Q^+ = Q^+_{\rm vis}
        = \frac{3}{8\pi} G \frac{M\dot{M}}{R^{3}}
        = 1.23 \times 10^{42}~m^{-2}~\dot{m}~r^{3}~\erg~\cm^{-2}~\s^{-1}.
\end{eqnarray}
The mass accretion rate $\dot M$ can be written in terms of the
viscosity coefficient $\nu$ as
\begin{eqnarray}
    \label{eq:mdot2}
    \dot{M} = 4\pi \rho R H v_{r} \approx 6\pi \nu \rho H,
\end{eqnarray}
where $v_{r}$ is the radial velocity of the gas.  For the viscosity,
we use the standard $\alpha$ prescription,
\begin{eqnarray}
    \label{eq:kinetic_visc}
    \nu = \alpha c_{S}^{2}/ \Omega_{K} = 0.1 \alpha_{-1}
    c_{S}^{2}/\Omega_{K}, \qquad \alpha_{-1}\equiv \alpha/0.1.
\end{eqnarray}
For simplicity, we have ignored a boundary correction term in
Eq.~(\ref{eq:mdot2}) (see Frank et al. 1992).

\subsubsection{Cooling rate}
\label{subsubsec:cool}

The rate of loss of energy by cooling has four contributions:
\begin{eqnarray}
    \label{eq:cooling}
    Q^- = Q^-_{\rm rad} + Q^-_{\rm photodiss}+ Q^-_{\rm adv} + Q^-_{\nu},
\end{eqnarray}
where $Q^-_{\rm rad}$ is the radiative cooling rate, $Q^-_{\rm
photodiss}$ is the cooling rate by photodissociation of heavy nuclei,
$Q^-_{\rm adv}$ is the advective energy transport (Abramowicz et
al. 1988; Narayan \& Yi 1994), and $Q^-_{\nu}$ is the cooling rate due
to neutrino loss.

The radiative cooling rate is expressed by
\begin{eqnarray}
    \label{eq:cooling2}
    Q^-_{\rm rad} = \frac{g_* \sigma_{\rm s} T^4}{2 \tau_{\rm tot}},
\end{eqnarray}
where $\sigma_{\rm s}=\pi^2 k_{\rm B}^4/(60 \hbar^3 c^2)$ is the
Stefan-Boltzmann constant and $g_{*}$ is the number of degrees of
freedom (= 2 for photons). The optical depth, $\tau_{\rm tot}$, is
given by
\begin{eqnarray}
    \label{eq:optical_depth}
    \tau_{\rm tot} = \kappa_{\rm R}\rho H = \frac{\kappa_{\rm R}\Sigma}{2},
\end{eqnarray}
where $\kappa_{\rm R}$ is the Rosseland-mean opacity,
\begin{eqnarray}
    \label{eq:opacity}
    \kappa_{\rm R} = 0.40 + 0.64 \times 10^{23}
    \left(\frac{\rho}{\g~\cm^{-3}}\right)\left(\frac{T}{\K}\right)^{-3}
    \g^{-1}~ \cm^2.
\end{eqnarray}
The first term on the right is from electron scattering and the second
is from free-free absorption.  The radiative optical depth in an NDAF
is extremely large; therefore, radiative cooling is negligible
compared to the other cooling terms described below (i.e., the flow is
extremely advection-dominated as far as the radiation is concerned).

Since the photodissociation of heavy nuclei requires energy, this
process acts like a cooling mechanism. The cooling rate is given by
\begin{eqnarray}
    \label{eq:photodiss}
    Q^{-}_{\rm photodiss} = q^-_{\rm photodiss}~H,
\end{eqnarray}
with
\begin{eqnarray}
    \label{eq:qdot_photodiss}
    q^-_{\rm photodiss} = 6.8 \times 10^{28}~\erg~\cm^{-3}~\s^{-1}
    \rho_{10} \left( \frac{A}{4} \right)^{-1} \left(\frac{B}{28.3\mev}
    \right) \left(\frac{v_{r}}{\cm~s^{-1}} \right) \left(\frac{dX_{\rm
    nuc}/dr}{\cm^{-1}} \right),
\end{eqnarray}
where $B$ is the binding energy of the nucleus (=28.3 MeV for
$^{4}$He), $A$ is the mass number of the nucleus, and $X_{\rm nuc}$ is
the mass fraction of nucleons, given by (Woosley \& Baron 1992; Qian
\& Woosley 1996)
\begin{eqnarray}
    \label{eq:Xnuc}
    X_{\rm nuc} = 295.5 \rho_{10}^{-3/4} T_{11}^{9/8}
    \exp\left(-0.8209/T_{11} \right).
\end{eqnarray}
In the context of an accretion disk, the radial velocity is
(see Eq.~(\ref{eq:mdot2}))
\begin{eqnarray}
    \label{eq:vr}
    v_{r} = \dot{M}/(2\pi R \Sigma) = 3\nu/(2R).
\end{eqnarray}
In Eq.~(\ref{eq:qdot_photodiss}), we have assumed that all the heavy
nuclei are $\alpha$'s.  For other nuclei, we only need to change the
binding energy $B$ and the mass number $A$ appropriately.

The advective cooling rate is given by (Kato, Fukue, \& Mineshige
1998),
\begin{eqnarray}
    \label{eq:advective_cooling}
    Q^-_{\rm adv}=\Sigma T v_r \frac{d s}{dR},
\end{eqnarray}
where $s$ denotes the entropy per particle,
\begin{eqnarray}
    \label{eq:entropy_per_matter}
    s = \left(s_{\rm rad} + s_{\rm gas}\right)/\rho.
\end{eqnarray}
Here, the entropy density of the radiation is
\begin{eqnarray}
    \label{eq:radiation-entropy}
    s_{\rm rad} = \frac{2}{3} a g_* T^3,
\end{eqnarray}
and the entropy density of the gas (i.e., nonrelativistic particles)
is
\begin{eqnarray}
    \label{eq:gas-entropy}
    s_{\rm gas} = \sum_i n_i\left(\frac52 +
      \ln\left[\frac{\g_i}{n_i}\left(\frac{m_i
          T}{2\pi}\right)^{3/2}\right]\right),
\end{eqnarray}
where the suffix $i$ runs over nonrelativistic nucleons and electrons,
and $g_i$ is the statistical degree of freedom of species $i$.  Note
that the entropy of degenerate particles is quite small. Therefore, we
neglect it.

It is convenient to define an advection parameter $f_{\rm adv}$ (e.g.,
Narayan \& Yi 1994),
\begin{eqnarray}
    \label{eq:fadv} f_{\rm adv} = {Q_{\rm adv}^{-}\over Q^{+}} \approx
    \left({H\over R} \right)^{2},
\end{eqnarray}
which measures the relative importance of advection.

\subsubsection{Neutrino emission}
\label{subsec:nucool}

Cooling by neutrino emission plays an important role in an NDAF and
requires careful discussion. When the accreting gas is transparent to
neutrinos, it is relatively straightforward to calculate the cooling
rate (see Narayan, Piran \& Kumar 2001; Kohri \& Mineshige
2002). However, once the gas becomes opaque to neutrinos, the cooling
rate is significantly modified.  To handle this regime, we introduce
the ``optical depth'' $\tau_{\nu_{i}}$ for each neutrino species,
$\nu_{i} = \nu_{e}$, $\nu_{\mu}$ or $\nu_{\tau}$, and follow the
approach of Di Matteo, Perna \& Narayan (2002) for estimating these
optical depths.\footnote{As per the approximations in Di Matteo, Perna
\& Narayan (2002), for simplicity, the difference of the optical
depths between $\nu_{e}$ and $\anti{\nu}_{e}$ is ignored. This
simplification does not change our conclusion so much.}

In the transparent limit, the total neutrino-cooling rate is simply
the sum of four terms, $(q^-_{Ne} + q^-_{e^+e^-} + q^-_{\rm brems} +
q^-_{\rm plasmon}) H$,
%
%
where $q^-_{Ne}$ is the cooling rate due to electron-positron capture
by a nucleon ``$N$'' ($=p, n$), $q^-_{e^+e^-}$ is from
electron-positron pair annihilation into neutrinos, $q^-_{\rm brems}$
is the cooling rate by nucleon-nucleon bremsstrahlung, and $q^-_{\rm
plasmon}$ is the cooling rate by plasmon decays (Kohri \& Mineshige
2002; Di Matteo, Perna \& Narayan 2002). In the following expressions,
we omit the Fermi-blocking effect by the background neutrinos in the
final state. We discuss this point later.

The electron-positron capture rate by nucleons is represented by the
sum of two terms:
\begin{eqnarray}
    \label{eq:e-capture}
    q^-_{Ne} = q^-_{p+e^- \to n + \nu_e} 
    + q^-_{n+e^+ \to p + \anti{\nu}_e},
\end{eqnarray}
with
\begin{eqnarray}
    \label{eq:e-p}
    q^-_{p+e^- \to n +
    \nu_e}\hspace{-0.3cm}&=&\hspace{-0.3cm}\frac{G_{\rm F}^2}{2\pi^3
    \hbar^3c^2} (1 + 3 g_A) n_p \int_Q^{\infty} \hspace{-0.3cm}dE_e
    E_e \sqrt{E_e^2 -
    m_e^2c^4}
    \left(E_e - Q\right)^3 \frac1{e^{\left(E_e - \mu_e\right)/k_{\rm
    B}T}+1},
\end{eqnarray}
\begin{eqnarray}
    \label{eq:e+n}
    q^-_{n+e^+ \to p +
    \anti{\nu}_e}\hspace{-0.3cm}&=&\hspace{-0.3cm}\frac{G_{\rm
    F}^2}{2\pi^3
    \hbar^3c^2} (1 + 3 g_A) n_n \int_{m_ec^2}^{\infty}
    \hspace{-0.3cm}dE_e E_e
    \sqrt{E_e^2 - m_e^2c^4} \left(E_e + Q\right)^3 \frac1{e^{\left(E_e
      + \mu_e\right)/k_{\rm B}T}+1},
\end{eqnarray}
where $G_{\rm F} = 2.302 \times 10^{-22} \cm~\mev^{-1}$, and the axial
vector coupling constant $g_A \simeq 1.4$, determined by the
experimental value of the neutron lifetime (see
\S~\ref{subsubsec:n-p-ratio}). We plot the two cooling terms in the
$T$--$\eta_{e}$ plane in Fig.~\ref{fig:amma_q_panel}~(c) and
Fig.~\ref{fig:amma_q_panel}~(d), respectively.  The rates are
normalized by the number density of the nucleon (i.e., proton or
neutron) in the initial state.

The electron-positron pair annihilation rate into neutrinos is the sum
of the contributions from the three lepton generations:
\begin{eqnarray}
    \label{eq:eeannihilation}
    q^-_{e^+e^-}=\sum_{i=e,\mu,\tau}
    q^-_{e^+e^-\to\nu_{i}\anti{\nu}_{i}},
\end{eqnarray}
with
\begin{eqnarray}
    \label{eq:eeannihilation1}
    q^-_{e^+e^- \to \nu_{e}\anti{\nu}_{e}}=
    3.4\times 10^{33}~\erg~ \cm^{-3}\s^{-1}
    \left(\frac{T}{10^{11}\K}\right)^9,
\end{eqnarray}
\begin{eqnarray}
    \label{eq:eeannihilation2}
    q^-_{e^+e^- \to \nu_{\mu}\anti{\nu}_{\mu}}=
    q^-_{e^+e^- \to \nu_{\tau}\anti{\nu}_{\tau}}=
    0.7\times 10^{33} \erg~ \cm^{-3}\s^{-1}
    \left(\frac{T}{10^{11}\K}\right)^9.
\end{eqnarray}
These expressions are valid in the nondegenerate limit $\eta_{e} \ll
1$. If the electrons are degenerate, the electron-positron pair
annihilation rate becomes quite small, compared with the other
neutrino cooling processes.  We therefore neglect the pair
annihilation term whenever $\eta_{e} \ll 1$.

The nucleon-nucleon bremsstrahlung rate through the process $ n + n
\to n + n + \nu + \anti{\nu}$ is represented by
\begin{eqnarray}
    \label{eq:brems_non-degN}
    q^-_{\rm brems} = 1.5 \times 10^{33} \erg~\cm^{-3}~\s^{-1}
    \left(\frac{T}{10^{11}\K}\right)^{5.5}\left(\frac{\rho}{10^{13}
      g~\cm^{-3}}\right)^{2},
\end{eqnarray}
where we consider only the case when the nucleons are not degenerate
(Hannestad \& Raffelt 1998; Burrows et al. 2000).

Plasmon decay into neutrinos is most effective at high density and
high electron degeneracy (Schinder et al. 1987).  The plasmons
$\tilde{\gamma}$ are photons interacting with electrons. The decay
rate into $\nu_e$ and $\anti{\nu}_e$ of transverse plasmons is given
by
\begin{eqnarray}
    \label{eq:plasmon}
    q^-_{\rm plasmon} = 1.5 \times 10^{32}
    \erg~\cm^{-3}~\s^{-1}\left(\frac{T}{10^{11}\K}\right)^{9}
    \gamma_p^6 e^{-\gamma_p}(1 + \gamma_p) \left(2 +
      \frac{\gamma_p^2}{1 + \gamma_p}\right),
\end{eqnarray}
where $\gamma_p = 5.565 \times 10^{-2} \sqrt{(\pi^2 + 3 \eta_e^2)/3}$
(Ruffert, Janka \& Sch\"afer 1996).  Note that the process
$\tilde{\gamma} \to \nu_e + \anti{\nu}_e$ is more effective by a
factor of $\sim 163$, compared with that of the other flavors, $\to
\nu_{\mu}\anti{\nu}_{\mu}$ and $\nu_{\tau}\anti{\nu}_{\tau}$ .

Having discussed the various cooling terms, we now introduce various
``optical depths'' of the different neutrinos. The absorption optical
depths of the three neutrino species $\nu_{i}=\nu_{e}, \nu_{\mu}$, and
$\nu_{\tau}$ are defined by
\begin{eqnarray}
    \label{eq:tauae}
    \tau_{a,\nu_{e}} =  \frac{
    \left( q^-_{p+e^- \to n + \nu_e}+q^-_{e^+e^- \to
    \nu_{e}\anti{\nu}_{e}} + q^-_{\rm brems} + q^-_{\rm plasmon}
    \right) H}
    { \displaystyle{(7/2)  \sigma T^{4}}
    },
\end{eqnarray}
\begin{eqnarray}
    \label{eq:tauamu}
    \tau_{a,\nu_{\mu}} =  \tau_{a,\nu_{\tau}}=   \frac{
    \left( q^-_{e^+e^- \to
    \nu_{\mu}\anti{\nu}_{\mu}} + q^-_{\rm brems} \right) H}
    { \displaystyle{(7/2)  \sigma T^{4}}
    }.
\end{eqnarray}
According to Tubbs \& Schramm (1975), Shapiro \& Teukolsky (1983), and
Di Matteo, Perna \& Narayan (2002), the scattering optical depth of
neutrinos through elastic scattering off background nucleons, $\nu_{i}
+ \left\{ n~{\rm or }~p \right\} \to \nu_{i} + \left\{n~{\rm or }~p
\right\}$, is the same for all three species and is given by
\begin{eqnarray}
    \label{eq:tause_nuN}
    \tau_{s,\nu_{i}} = 7.7 \times 10^{-7} \left(C_{s,p}Y_{p} +
    C_{s,n}Y_{n} \right) T_{11}^{2} \rho_{10} H.
\end{eqnarray}
Here $C_{s,p} = [4(C_{V}-1)^{2}+5\alpha_{a}^{2}]/24$ and
$C_{s,n}=(1+5\alpha_{a}^{2})/24$, with the vector coupling $C_{V} =
1/2+2\sin^{2}\theta_{W}$, and $\alpha_{a} \approx 1.25$. The Weinberg
angle is $\sin^{2}\theta_{W} = 0.23$ (Eidelman et al. 2004).

Using the above optical depths of neutrinos, we write the total
neutrino cooling rate as
\begin{eqnarray}
    \label{eq:totnucool}
    Q_{\nu}^{-} &=& \sum_{i=e,\mu,\tau} \frac{(7/8) \sigma
    T^{4}}{(3/4)\left[ \tau_{\nu_i}/2 + 1/\sqrt{3} +
    1/(3\tau_{a,\nu_{i}}) \right] } \nonumber \\
    &=&  6.62 \times 10^{39} T_{11}^{4} \erg~\cm^{-3}~\s^{-1}
    \sum_{i=e,\mu,\tau}
    \frac{1}{\left[ \tau_{\nu_i}/2 + 1/\sqrt{3} +
    1/(3\tau_{a,\nu_{i}}) \right]},
\end{eqnarray}
where the total optical depth of the neutrino is $\tau_{\nu_i} =
\tau_{a,\nu_{i}} + \tau_{s,\nu_{i}}$.  The above expression is based
on the work of Popham \& Narayan (1995) and is designed to operate in
both the optically very thin and optically very thick limits.  It
represents a bridging formula between these two limits and provides a
reasonable estimate of the cooling rate in the difficult but important
intermediate regime where $\tau$ is of order a few.

Note that the optical depths discussed above provide some information
on whether or not neutrinos are thermalized in the electromagnetic
thermal bath (provided the timescale for collisions is shorter than
the dynamical timescale\footnote{We have checked the competition among
various timescales in Appendix~\ref{sec:timescale}. We find that the
timescale for neutrino collisions is much shorter than the dynamical
timescale in the parameter space where the optical depths are larger
than 2/3.}).  Because we do not explicitly solve the Boltzmann
equation for the time evolution of the energy distribution of
neutrinos and the energy transfer, we conservatively regard the above
neutrino optical depths as an indicator of the degree of
thermalization of neutrinos.  That is, when all the $\tau$'s are
larger than 2/3, we assume the neutrinos to be thermalized.

For the other scattering processes except neutrino-nucleon scattering
($\nu_{i} + \left\{ n~{\rm or }~p \right\} \to \nu_{i} + \left\{n~{\rm
or }~p \right\}$), there also exists the possibility of elastic
scattering off background electrons (positrons), $\nu + e \to \nu +
e$. The energy dependence (or temperature dependence) of this reaction
rate is similar to that of neutrino-nucleon collisions. However, its
contribution to the optical depth is subdominant or comparable at most
in the electron-degeneracy regime. Therefore, for simplicity we have
omitted it in the calculation of the neutrino optical
depth.\footnote{For energy transfer of neutrinos, however, the
scattering off background electrons might be more important than
scattering off nucleons. Thus, if we were to explicitly solve the
Boltzmann equation for the energy transfer of neutrinos, we would have
to include the elastic collisions between neutrinos and electrons.}

Finally we note that, when we calculate the neutrino-cooling rates, we
omit the Fermi-blocking effect by the background neutrinos.  We
believe that this approximation is reasonable.  When the accreting gas
is transparent to neutrinos ($\tau{\rm 's} \ll 1$), the approximation
is obviously good since the produced neutrinos are not thermalized,
and so we can omit the neutrino-neutrino scattering.  On the other
hand, when the gas becomes opaque to neutrinos, we strongly suppress
the neutrino-cooling rates by introducing the optical depths $\tau$ in
Eq.~(\ref{eq:totnucool}). Since the suppression effect due to optical
depth is typically much stronger than the suppression due to
Fermi-blocking for $\tau{\rm 's} \gg 1$, we believe it is reasonable
to ignore the latter.

\subsection{Comparison to previous work}
\label{subsec:newpoints}

Several aspects of our model are similar to previous treatments of the
problem (Narayan, Piran \& Kumar 2001; Kohri \& Mineshige 2002; Di
Matteo, Perna \& Narayan 2002), though our work goes significantly
further in handling the details.  The original model of Narayan et
al. (2001) made use of a simple neutrino-cooling rate, which was then
improved by Kohri \& Mineshige (2002) who pointed out that it is
important to include the effect of electron degeneracy which
suppresses the neutrino-cooling rate at high density and high
temperature. Di Matteo, Perna \& Narayan (2002) pointed out that the
accreting gas could be optically thick to neutrino emission, so that a
neutrinosphere could form at high temperatures and thereby suppress
neutrino emission.  The accretion flow then becomes
advection-dominated~\footnote{Recently, we noticed that Yokosawa,
Uematsu \& Abe (2004) studied models of the neutrino-cooled disk
without explicitly adopting the advective cooling term.  Their results
are different from those in Narayan, Piran \& Kumar (2001), Kohri \&
Mineshige (2002), and Di~Matteo, Perna \& Narayan (2002), as they have
admitted.}.

Our model includes the effects of electron degeneracy and neutrino
optical depth in calculating physical quantities such as
neutrino-cooling rates, matter density, temperature, pressure, etc.
In this connection, we have taken greater care to calculate the
neutron to proton ratio $n/p$ as accurately as possible since this
ratio has a large effect on the neutrino-cooling rates and the
electron chemical potential. Although the importance of $n/p$ was also
stressed by Kohri and Mineshige (2002), their approach was a great
deal simpler.  Here we have considered the competition among various
timescales in determining the equilibrium between neutrons and
protons, and we have also checked under what conditions neutrinos are
thermalized.

Furthermore, we have calculated various quantities such as the
neutrino cooling rates, the interconversion rates between neutrons and
protons, the electron pressure, and the electron number densities by
numerically integrating the distribution function of electrons over
momentum.  We are thus able to calculate these quantities even in the
delicate regime where the electron degeneracy is moderate. This is a
significant improvement over previous works which were restricted to
calculating quantities only in the two opposite limits of extremely
degenerate electrons and fully non-degenerate electrons.

\subsection{Results for the disk structure}
\label{sec:disk_structure}

Using the model described in the previous subsections, we can
calculate all the properties of a neutrino-cooled accretion disk for
any large mass accretion rate.  Given the mass $m$ of the accreting
compact core, the mass accretion rate $\dot{m}$, and the radius $r$
(all in dimensionless units, see Eqs.~(\ref{eq:smallm}),
(\ref{eq:smallmdot}) and (\ref{eq:smallr})), and using the various
subsidiary relations written down earlier, we numerically solve the
energy balance condition (heating rate = cooling rate),
\begin{equation}
Q^+ = Q^-.
\end{equation}
The solution gives the various quantities of interest such as the mass
density $\rho$, the temperature $T$, the surface density $\Sigma$,
etc.

\begin{figure}
\epsscale{1.0}
\plotone{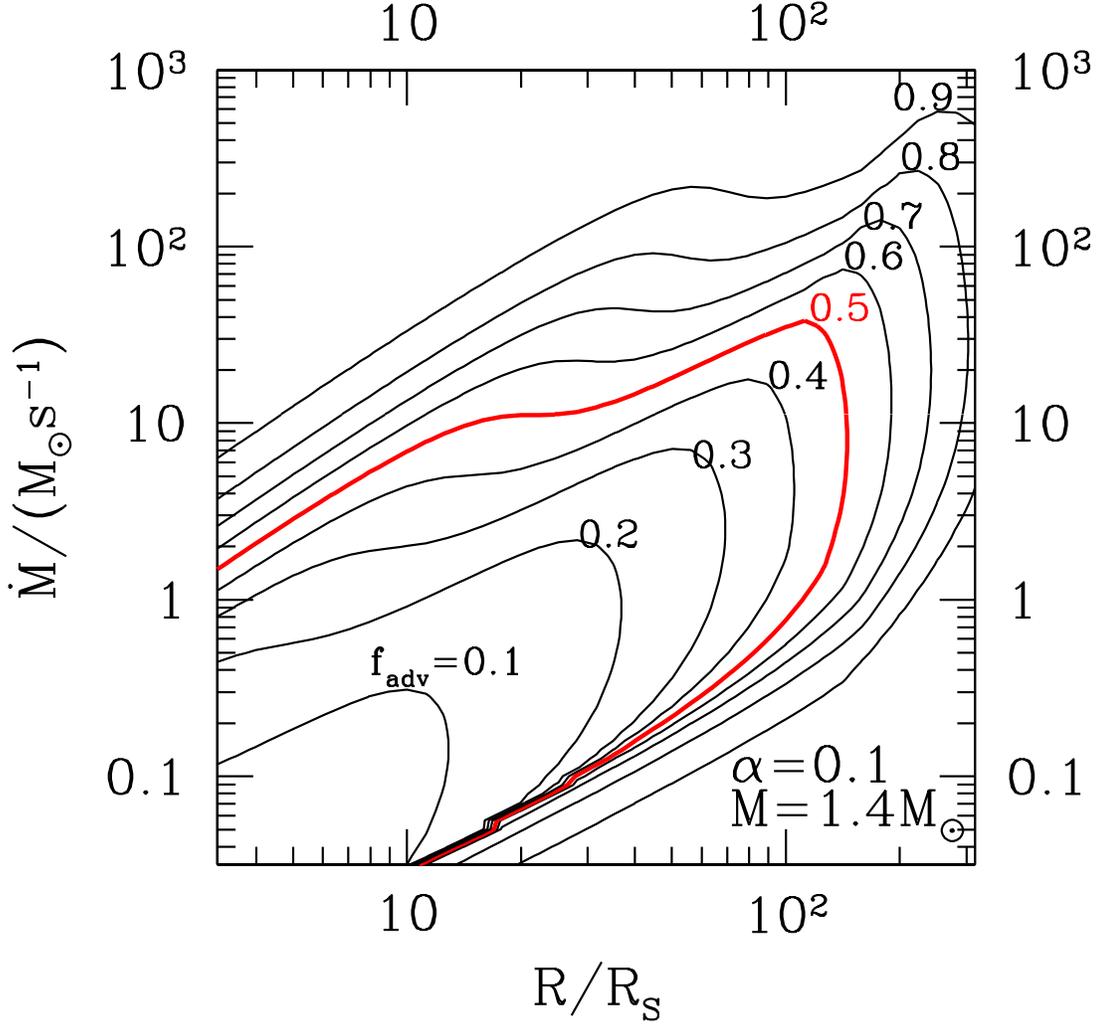}
\vspace{.0in}
\caption{Contours of the advection parameter $f_{\rm adv} \equiv
Q_{\rm adv}^-/Q^+$ in the $r$--$\dot{m}$ plane, where $r\equiv R/R_S$
and $\dot m \equiv \dot M/M_\odot{\rm s^{-1}}$. The parameter $f_{\rm
adv}$ is a measure of the disk thickness (see eq.~\ref{eq:fadv}) and
also determines how susceptible the disk is to producing an outflow
(\S\S~3.1, 3.2). The results shown correspond to a viscosity parameter
$\alpha = 0.1$ and a compact central object of mass $M=1.4
M_{\odot}$.}
\label{fig:f}
\end{figure}

Fig.~\ref{fig:f} shows contours of the advection parameter $f_{\rm
adv} \equiv (H/R)^2$ in the $r$--$\dot{m}$ plane for $m=1.4$
(corresponding to a proto-neutron star at the center) and $\alpha =
0.1$.  Note the qualitative similarity of this plot to the middle
panel of Fig.~3 in Di Matteo, Perna \& Narayan (2001).  However, the
present calculations are more accurate and also correspond to a
different mass (Di Matteo et al. considered $m=3$).  In this plot,
regions with $f_{\rm adv}$ close to unity are highly
advection-dominated, whereas regions with $f_{\rm adv} < 0.5$ are
cooling-dominated and have significant neutrino emission.

\begin{figure}
\epsscale{1.0}
\plotone{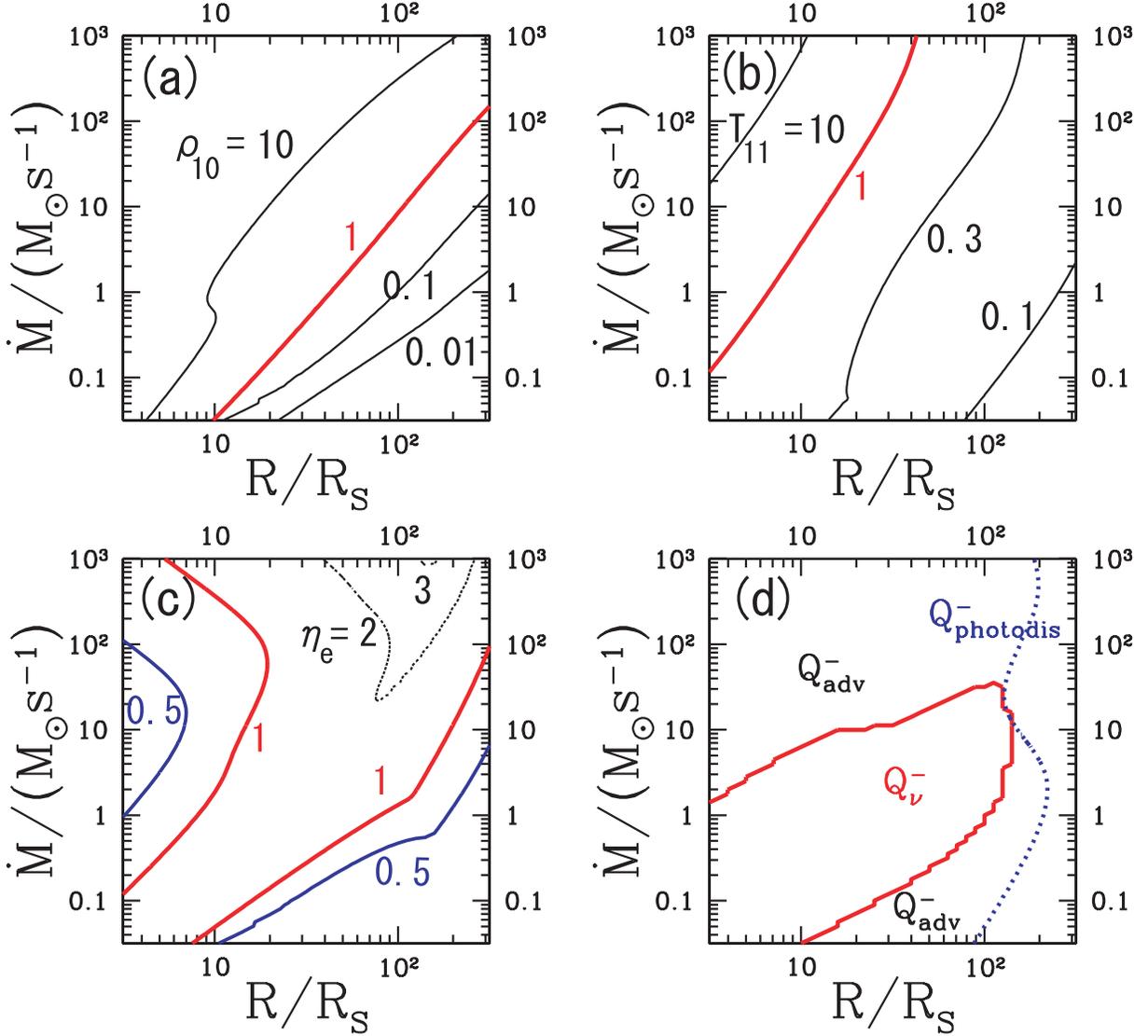}
\vspace{.0in}
\caption{ (a) Contours of the matter density $\rho_{10} = \rho/10^{10}
~\g\,\cm^{-3}$ in the $r$--$\dot{m}$ plane. The thick solid line
corresponds to $\rho_{10}=1$. (b) Contours of $T_{11} =
T/10^{11}~\K$. (c) Contours of $\eta_{e}$. Note that $\eta_{e}$
becomes large toward the upper right region.  (d) Dominant cooling
process in various regions of the $r$--$\dot{m}$ plane. In the central
region, neutrino cooling $Q^{-}_{\nu}$ dominates, with
electron-positron capture being the most important process.  In the
region on the right, photodissociation of nuclei is important.  The
dotted line shows the contour corresponding to mass fraction of
nucleons $X_{\rm nuc} = 0.5$, where most nuclei are destroyed by
endothermic photodissociation (see
Fig.~\ref{fig:xnuc_np_p_ne_panel}(a) and the discussion in
Section~\ref{subsubsec:cool}).  Advective cooling dominates in the
rest of the plane, as seen also in Fig.~\ref{fig:f}.  All panels
correspond to $\alpha = 0.1$ and $M=1.4 M_{\odot}$.}
\label{fig:rho_T_eta_q}
\end{figure}

Figs.~\ref{fig:rho_T_eta_q}~(a) and (b) show contours of density and
temperature, respectively, in the $r$--$\dot m$ plane.  Both
quantities increase toward the upper left region of the diagram. The
degeneracy parameter $\eta_{e}$ is shown in
Fig.~\ref{fig:rho_T_eta_q}~(c). There is a tendency for $\eta_{e}$ to
increase toward the upper right region.

In Fig.~\ref{fig:rho_T_eta_q}~(d), we show which of the various terms
in $Q^-$ dominates in which regions of the plane.  As already
mentioned, radiative cooling is never important at these ultra-high
mass accretion rates.  Neutrino cooling, on the other hand, does
become important over an extended region of the plane near the middle
of the plot. The most important emission process here is
electron-positron capture.  Not surprisingly, the region where
neutrino cooling dominates overlaps with the region where $f_{\rm
adv}$ is small in Fig. \ref{fig:f}, i.e., where advection is not important.

Figures \ref{fig:f} and \ref{fig:rho_T_eta_q}~(d) show that advection
is important over much of the rest of the plane.  The reason is easy
to understand.  Contours of $\tau_{a,\nu_{e}}$ and $\tau_{s,\nu_{e}}$
are shown in Fig.~\ref{fig:tau_nue_panel}~(a) and
Fig.~\ref{fig:tau_nue_panel}~(b), with the thick lines corresponding
to $\tau = 2/3$. From these plots, we see that the disk becomes
``optically thick'' to neutrino emission, and a neutrinosphere is
formed, in the upper left region of the plane.  This suppresses
neutrino cooling in this region of parameter space, causing the flow
to become advection-dominated.  In the bottom right region, the
optical depth is very small, but the emission processes themselves are
weak and so once again the flow is advection-dominated.

In the right-most region of Fig.~\ref{fig:rho_T_eta_q}~(d), cooling
due to photo-dissociation dominates.  This is indicated in
Fig.~\ref{fig:xnuc_np_p_ne_panel}~(a), which shows contours of the
nucleon faction $X_{\rm nuc}$ in the $r$--$\dot{m}$ plane. For $r
\lesssim 150$, we find that approximately $X_{\rm nuc} \sim
1$. Therefore, for these radii which are the parameter space of
interest to us, most of the nuclei have been completely dissociated
into free nucleons in the thermal bath.  We may omit (for simplicity)
the cooling due to photodissociation of heavy nuclei in
Eq.~(\ref{eq:qdot_photodiss}) in this region of the plane. At larger
radii, photodissociation is important and dominates the cooling.

In Fig.~\ref{fig:xnuc_np_p_ne_panel}~(b) we show the neutron to proton
ratio.  We see that $n/p$ is larger than unity over much of the
parameter space. This is because electrons are highly degenerate and
have positive values of $\eta_{e}$.  We then expect an excess of
neutrons.  This neutron rich gas with a short dynamical timescale
($\ll 1~\s$) might lead to $r$-process nucleosynthesis (Hoffman,
Woosley \& Qian 1997), as we discuss in \S~\ref{subsec:nuclear}.  In
Fig.~\ref{fig:xnuc_np_p_ne_panel}~(c) the dominant component of the
pressure (see \S~\ref{subsec:press} for a definition of the various
components) is shown in the $r$--$\dot{m}$ plane.  We see that gas
pressure dominates in the region where the neutrino cooling is
effective. Finally, in Fig.~\ref{fig:xnuc_np_p_ne_panel}~(d) we plot
contours of the net number density of electrons $n_e$.

\section{Outflow Energy From the Disk}
\label{sec:outflow}

In this section we present several different estimates of the
mechanical energy in the wind flowing out from a neutrino-dominated
disk during core-collapse of a massive star.  We begin in
\S~\ref{sec:qualitative} with a simple discussion of the relevant
physics and follow this up in succeeding subsections with more
detailed calculations.

\subsection{Qualitative Estimate of Outflow Energy}
\label{sec:qualitative}

Before going into detailed calculations, we first present a simple
qualitative estimate of the energy that might be carried out by the
disk outflow.  A vigorous outflow is expected whenever the accretion
flow is advection-dominated (Narayan \& Yi 1994, 1995a).  Although the
calculations in \S~\ref{sec:disk_structure} and the results shown in
Fig.~\ref{fig:f} indicate that advection does not dominate for all
parameters of interest, let us for simplicity assume here that the
disk is always advection-dominated.  Since an outflow carries away
mass, the accretion rate in the disk will decrease with decreasing
radius.  For simplicity, let us assume that the accretion rate varies
as a power-law in radius,
\begin{eqnarray}
    \label{eq:Mdot_simple}
    \dot{M}(r) = \dot{M}_{0} \left( \frac{r}{r_{0}}  \right)^{s},
\end{eqnarray}
with a constant index $s$.  Here $r_0$ is the radius of the outer edge
of the disk and $\dot M_0$ is the mass accretion rate at that
radius. The differential rate of outflow of mass in the wind is then
given by
\begin{eqnarray}
    \label{eq:dmdot}
    d \dot{M} = \frac{s \dot{M}_{0}}{r_{0}^{s}} \frac{dr}{r^{1-s}}.
\end{eqnarray}

The terminal velocity of the outflowing gas is likely to be of the
order of the escape velocity from the point of origin in the disk,
$v_{\rm esc} = \sqrt{GM/R}=c/\sqrt{2r}$.  Let us therefore write the
specific energy (at infinity) of the wind as $(1/2)\xi v_{\rm
esc}^{2}$, where the fudge factor $\xi \sim 0.1-1$ absorbs our
ignorance of the details of the outflow.  The differential rate of
outflow of energy in the wind is then
\begin{eqnarray}
    \label{eq:dEdot}
    d \dot{E}_{w} = \xi \frac{GM}{R} d \dot{M} = \frac{s}{2} \frac{\xi
    \dot{M}_{0} c^{2}} {r_{0}^{s}} \frac{dr} {r^{2 - s}}.
\end{eqnarray}
Let us assume that the scaling of $\dot{M}$ given in
Eq.~(\ref{eq:Mdot_simple}) extends from $r=r_{0}$ on the outside down
to an innermost radius $r= r_{\rm in}$. Integrating over $r$, we then
estimate the total rate of outflow of energy in the wind to be
\begin{eqnarray}
    \label{eq:totE2}
    \dot{E}_{w} =
\left\{
 \begin{array}{ll}
    \displaystyle{
     \frac{s}{2(1-s)} \frac{\xi \dot{M}_{0} c^{2}}{r_{0}^{s}}
     \left(\frac1{r_{\rm in}^{1-s}} - \frac1{r_{0}^{1-s}} \right)
    }, \qquad  {\rm for}~s < 1,
\\
\\
    \displaystyle{
     \frac{s}{2} \frac{\xi \dot{M}_{0} c^{2}}{r_{0}}
     \ln\left(\frac{r_{0}}{r_{\rm in}} \right)
     }, \qquad  {\rm for}~s = 1.
\\
 \end{array}
 \right.
\end{eqnarray}
If the compact core at the center is a non-spinning black hole, then
we expect $r_{\rm in}=3$ (innermost stable circular orbit in a
Schwarzschild space-time), and if it is a spinning hole then $r_{\rm
in}$ would be even smaller.  For a proto-neutron star, however, the
inner edge of the accretion flow will be significantly larger.  If the
mass of the core is $1.4M_\odot$ and its radius is about 30 km (e.g.,
Lattimer \& Prakash 2004), then we have $r_{\rm in}\sim7$, which is
the value we assume below.

Consider first the second line of equation~(\ref{eq:totE2}) which
corresponds to the limit $s=1$.  We see that the outflow energy
decreases with increasing outer radius.  This is because $s=1$
corresponds to a very heavy mass outflow rate; in fact, it is the
maximum likely value of $s$ (see Blandford \& Begelman 1999). Even
though the specific energy of the outflowing matter increases at
smaller radius, the amount of mass available at small radii is greatly
reduced. Therefore, the overall scale of the energy outflow is
determined primarily by the outer radius, and we get equal
contributions to the outflow energy from equal logarithmic intervals
of radius, hence the logarithmic factor.  Figure~\ref{fig:Edot0} shows
a plot of $\dot E_w$ versus $r_0$.  We see that for $s=1$ the maximum
energy outflow rate is obtained for $r_0\sim15$. At smaller radii,
even though the overall energy scale is larger, the value of the
logarithm becomes small and this causes a reduction in the outflow
energy.

In the opposite limit, when $s$ is close to 0, we see from the first
line of equation~(\ref{eq:totE2}) that the energy outflow rate is
proportional to $s\dot M_0/r_{\rm in}$.  Here, the rate is almost
independent of $r_0$ because the energy outflow is dominated by small
radii.  In fact, what matters now is how much mass is expelled in the
outflow near $r_{\rm in}$, which by equation~(\ref{eq:dmdot}) is
proportional to $s$. When $s$ is exactly equal to 0, there is
obviously no outflow at all and hence the outflow energy vanishes.
Figure~\ref{fig:Edot0} shows the behavior of the outflow energy for
intermediate values of $s$ between 0 and 1.  The overall pattern is
easy to understand in terms of the two limiting cases discussed above.

Quantitatively, Fig.~\ref{fig:Edot0} shows that $\dot E_w/ \xi\dot
M_0c^2$ is in the range 0.01 to 0.05 for reasonable choices of $r_0$
and $s$.  In other words, for every solar mass supplied to the disk at
the outer edge $r_0$, an amount of energy $\sim (2-10)\times
10^{52}\xi$ erg is expected to flow out of the disk in the wind. Even
if we conservatively take $\xi$ to be 0.1, this is still a substantial
amount of energy.  We have, however, assumed in the present discussion
that the disk is highly advection-dominated throughout and that it
ejects a strong outflow for all combinations of $\dot M$ and $R$. In
reality, the accretion flow in an NDAF is only partially
advection-dominated and the degree of advection varies as a function
of mass accretion rate and radius.  In the following subsections we
obtain better estimates of the energy in the outflow.

\subsection{More Quantitative Results}
\label{subsec:generic}

To allow for the effect of variable advection, let us generalize
equation~(\ref{eq:dmdot}) to
\begin{eqnarray}
    \label{eq:lnMlnR} \frac{d\ln \dot{M}}{d \ln R} = s(R) \geq 0,
\end{eqnarray}
i.e.,
\begin{eqnarray}
    \label{eq:lnMlnR2}
    \frac{d \dot{M}}{d R} = s(R) \frac{\dot{M}}{R},
\end{eqnarray}
where the index $s$ is now no longer considered to be a constant but
is allowed to vary with radius.  As already mentioned, we expect
outflows to be important when the accreting gas is advection-dominated
and to be negligible when the gas is able to cool readily.  Based on
this insight, Yuan, Cui \& Narayan (2004) came up with the following
simple prescription for the outflow index:
\begin{eqnarray}
    \label{eq:S_R}
    s(R) = s_{0} f(R)= s_{0} f_{\rm adv},
\end{eqnarray}
where $s_0$ is a constant, and $f_{\rm adv} = Q_{\rm adv}^{-}/Q^{+}
\approx (H/R)^{2}$ measures the degree of advection in the flow (see
eq.~(\ref{eq:fadv})).  Although Yuan et al.  (2004) proposed this
model for advection-dominated flows in X-ray binaries, the arguments
behind it are general and should apply also to an NDAF.

The differential outflow energy flux produced between radii $r$ and
$r+dr$ is
\begin{eqnarray}
    \label{eq:outflow_rate}
    d\dot{E}_{w} =  \frac12 \xi v_{\rm esc}^{2} d\dot{M} =
    s(r)\frac{\dot{M}}{r} \frac{\xi c^{2}}{4 r} dr,
\end{eqnarray}
where $\xi$ is the same fudge factor introduced in
\S~\ref{sec:qualitative}.  Integrating from $r= r_0$ down to $r=r_{\rm
in}$ we obtain the total rate of energy outflow from the disk
$\dot{E}_{w}$.  In Fig.~\ref{fig:Edotcontour_panel}, we show contours
of $\dot{E_{w}}/(\xi \dot{M}_{0}c^{2})$ in the $r_{0}$--$\dot{m}_{0}$
plane (where $\dot m_0 = \dot M_0/M_\odot{\rm s^{-1}}$) for different
choices of the outflow index: $s_{0}=$ 0.3, 0.5, 0.7, and 0.9. For a
wide range of $r_0$ and $\dot{M}_{0}$, we see that roughly
$(0.1-1)\xi$~\% of the rest mass energy of the accreting gas comes out
in the form of kinetic energy in the wind. The energy estimate is
somewhat lower than the one given in \S~\ref{sec:qualitative}, mainly
because the disk is only moderately advection-dominated over wide
ranges of the parameters.  Nevertheless, if $\dot{M}_{0}$ = $1
M_{\odot}~\s^{-1}$ ($0.1 M_{\odot}~\s^{-1}$), then $\dot{E}_{w}$ is
$(10^{51} - 10^{52})\xi$ erg~s$^{-1}$ [$(10^{50} - 10^{51})\xi$
erg~s$^{-1}$]. Thus, even for fairly small values of $\dot{M}_{0}$, we
find that the outflow energy is still quite significant.

What value of $s_{0}$ should we use for the outflow?  The limit
$s_{0}=0$ corresponds to the original ADAF (Narayan \& Yi 1994; 1995a)
in which there is no outflow, while $s_{0}=1$ corresponds to the
extreme limit of a convection-dominated accretion flow (CDAF; Narayan,
Igumenshchev \& Abramowicz 2000; Quataert \& Gruzinov 2000) which
deviates maximally from the ADAF.  Values of $s_0$ in between
correspond to generalized wind models (Blandford \& Begelman
1999). Numerical hydro and MHD simulations give $s_{0}$ anywhere in
the range from about 0.7 to 1 (Igumenshchev et al. 2000; 2003; Pen et
al., 2003). On the other hand, a model for the accreting supermassive
black hole Sgr A* at the Galactic Center gives $s_{0} \sim 0.3$ (Yuan
et al. 2003).  It is probably best to keep an open mind on the value
of $s_{0}$, though larger values should perhaps be favored.

The results shown in Figure \ref{fig:Edotcontour_panel} correspond to
a viscosity parameter $\alpha=0.1$. While this is a reasonable value,
it is of interest to investigate how the results change for other
values.  Figure~\ref{fig:Edotcontour_panel_a0.01} shows the effect of
changing $\alpha$ to 0.01.  We see that the contours in the lower half
of the various panels are shifted downward by about two orders of
magnitude in $\dot M_0$. (Note that the vertical axis extends over a
larger range in Figure~\ref{fig:Edotcontour_panel_a0.01}.)  The effect
is exactly the same as for ADAFs with {\it radiative cooling}.  As
explained in Narayan \& Yi (1995b; see also Narayan, Mahadevan \&
Quataert 1998), at low Eddington ratios, the transition from a
radiatively efficient accretion flow to an advection-dominated
accretion flow (ADAF) occurs at a critical mass accretion rate $\dot
M_{\rm crit} \propto \alpha^2$.  This scaling arises because the
cooling rate per unit mass of the accreting gas is proportional to
$\rho$, which is proportional to $\dot M\alpha^{-1}$, so that the
cooling time of the gas varies as $t_{\rm cool} \propto \alpha/\dot
M$.  On the other hand, the accretion time varies as $t_{\rm acc}
\propto \alpha^{-1}$. Thus, the condition $t_{\rm cool} = t_{\rm
acc}$, which represents the transition from a radiatively efficient
flow to an ADAF, leads to $\dot M_{\rm crit} \propto\alpha^{2}$.  In
an NDAF, the cooling is via neutrinos.  However, the cooling rate per
unit mass is still proportional to the density and so the same scaling
continues to hold.

\subsection{Application to Supernovae --- Prompt Explosion}
\label{subsec:application1}

In this and the following subsection, we consider two distinct
(idealized) scenarios in which the outflow from an NDAF might cause a
supernova explosion. Here we suppose that the collapsing stellar core
has a fairly large specific angular momentum so that immediately after
the initial homologous collapse, we have a proto-neutron star core
plus some additional material in a surrounding accretion disk.  The
disk forms nearly on a dynamical time, which is much shorter than the
viscous time of the orbiting gas.  Thus, we have a fully formed disk
at a particular instant $t=0$ and we follow the depletion of the mass
in the disk as a result of viscous accretion and outflow.  The
question we are interested in is how much energy flows out in the wind
and whether this energy is enough to cause a successful prompt
supernova explosion.

Let us assume that the disk initially consists of a mass
\begin{eqnarray}
    \label{eq:M_d_0}
    M_d(0)  \equiv~M_d(t=0),
\end{eqnarray}
with the bulk of the mass lying at some characteristic radius
$R_0=r_0R_S$, which we will think of as the ``outer radius'' of the
disk.  For clarity, in this subsection we explicitly write the
argument of functions such as $t$ and $R$. From equations
(\ref{eq:mdot2}) and (\ref{eq:kinetic_visc}), we see that the viscous
timescale at radius $R$ is given by
\begin{equation}
    t_{\rm visc}(t,R) = {R\over v_R(t,R)} \sim
    {2\over3\alpha}{R^2\Omega_K(R) \over \left[c_s(t,R)\right]^2}.
\end{equation}
The mass accretion rate at the outer radius $R_{0}$ is then
approximately given by
\begin{equation}
    \label{eq:dMd_dt}
    \dot{M}_{0}(t) \equiv \dot{M}(t,R=R_{0})= - dM_d(t)/dt \sim
    M_d(t)/t_{\rm visc}(t,R_0),
\end{equation}
where $\dot{M}(t,R)$ refers to the mass accretion rate at a given time
$t$ and at a radius $R$.  The above equation gives both the mass
accretion rate at the outer edge at a given time and the rate at which
the total disk mass decreases.  In addition, at a given time, $\dot M
(t,R)$ decreases with decreasing $R$ according to equations
(\ref{eq:lnMlnR2}) and (\ref{eq:S_R}) written down earlier.  Finally,
we need a prescription for the evolution of the characteristic outer
radius of the disk.  For simplicity, we assume that the radius does
not change with time. We then have a complete prescription for the
temporal and radial variation of the accretion rate $\dot{M}(t,R)$.
The time evolution of the disk mass $M_{d}(t)$ is obtained by solving
Eq.~(\ref{eq:dMd_dt}),
\begin{eqnarray}
    \label{eq:Md}
    M_{d}(t) = M_{d}(0) \exp\left[- \int_{0}^{t}d\tau \frac1 {t_{\rm
    visc} (\tau ,R_0)} \right].
\end{eqnarray}
Equivalently, the time evolution of the outer mass accretion rate
$\dot M_0(t) = \dot M (t,R_0)$ is given by
\begin{eqnarray}
    \label{eq:Mdot_t_R}
     \dot{M}_{0}(t) = \dot{M}_{0}(0)
     \frac{t_{\rm visc} (0,R_0) } {t_{\rm visc} (t,R_0)} \exp\left[-
     \int_{0}^{t}d\tau \frac1 {t_{\rm
    visc} (\tau ,R_0)} \right]
\end{eqnarray}
With these formulae, for given initial disk mass $M_d(0)$, outer
radius $r_0=R_0/R_S$ and outflow index $s_0$, we can calculate the
complete evolution of the disk and estimate the total energy,
integrated over radius and time, carried out by the outflow.
Figure~\ref{fig:tinteg_prompt_panel} shows some numerical results.

In Fig.~\ref{fig:tinteg_prompt_panel}(a) we plot contours of the total
outflow energy $E_{w}/\xi$ in the $r_{0}-M_{d}(0)$ plane for
$s_{0}=0.5$.  For any value of $\xi \ne 1$, the outflow energy is
obtained by taking the value given in the plot and multiplying by
$\xi$. Note that $E_{w}/\xi$ exceeds $10^{50}$ erg over a wide region
of parameter space, in particular for $M_{d}(0) \gtrsim 10^{-2}
M_{\odot}$ and $r_{0} \sim 10 - 100$.  In
Fig.~\ref{fig:tinteg_prompt_panel}(b), we show $E_{w}/\xi$ as a
function of $s_{0}$ for selected values of the disk mass:
$M_{d}(0)/M_{\odot}$ = $10^{-2}$, $5 \times 10^{-2}$, $0.1$, 0.5, and
1. Here we have set $r_{0}=20$. Note again that the outflow energy is
fairly substantial, especially if the disk mass is $0.1M_\odot$ or
more.  These results show that the outflow can help to produce a
prompt supernova explosion provided there is enough mass in the
initial accretion disk.

\subsection{Application to Supernovae --- Delayed Explosion}
\label{subsec:application2}

In this subsection we consider a different scenario. Let us suppose
that the initial collapse does not lead to a prompt explosion, either
because there is not enough angular momentum in the core to form an
accretion disk or because there is insufficient mass in the disk.  The
system will transition to a fairly long-lived state (several seconds)
in which material will flow in through the stalled shock towards the
central core.  As time progresses, the infalling material will
originate from larger and larger radii in the pre-collapse star and it
is quite likely that the angular momentum of the gas will be
sufficient to produce a centrifugally-supported fallback disk.  For a
core of mass $1.4M_\odot$ and a radius of 30 km, the critical specific
angular momentum needed to form a disk is $\ell_{\rm crit}\sim
2.5\times10^{16} ~{\rm cm^2\,s^{-1}}$.  This level of angular momentum
is not unusual for a collapsing star (e.g., Mineshige et al. 1997;
Heger, Langer \& Woosley 2000).

{}From a computational point of view, the present scenario differs from
the one described in the previous subsection primarily in the relative
magnitude of time scales. Previously, we assumed that the fallback
time over which matter is added to the disk $t_{\rm fallback}$ is much
shorter than the viscous timescale $t_{\rm visc}$ over which a given
parcel of gas flows through the disk, i.e., $t_{\rm fallback} \ll
t_{\rm visc}$. Thus, the time evolution of the disk is primarily
determined by $t_{\rm visc}$.  In contrast, here we assume that the
viscous timescale is much shorter than the fallback time: $t_{\rm
visc} \ll t_{\rm fallback}$. Thus, the time evolution is determined by
$t_{\rm fallback}$.

Assuming that a disk forms during the fallback stage, it is clear that
a considerable amount of outflow energy will be generated in the
present scenario.  This is because a failed supernova has few to
several solar masses of fallback material. Even if only a fraction of
this mass goes into the disk, the outflow energy would still be
considerable.  To get some numerical estimates, let us assume that the
fallback material rains down at some ``outer'' radius $R_0$ at an
initial rate $\dot M_0(0)$. Let us also assume that the mass inflow
rate declines exponentially with time with a characteristic decay time
$\bar{t}$,
\begin{eqnarray}
    \label{eq:M_0_t}
    \dot{M}_{0}(t) = \dot{M}_{0}(0) \exp\left(-t/\bar{t}
    \right), \qquad {\rm for}~t \ge 0,
\end{eqnarray}
Note that the subscript ``0'' means the value at $R=R_{0}$. The
variation of $\dot M$ with $R$ is as in equations (\ref{eq:lnMlnR2})
and (\ref{eq:S_R}).  In this simple model, the total amount of mass
flowing into the disk at $R_0$ is $\bar{t}\dot M_0(0)$.

In Fig.~\ref{fig:tinteg_panel}~(a), we plot contours of the time- and
radius-integrated total outflow energy as a function of the initial
accretion rate $\dot M_0(0)$ and the outer radius $r_0$.  For
simplicity, we have assumed that $r_{0}$ is independent of time (see
below for a more realistic model), and we have adopted $\bar{t} =
1~\s$, $s_{0}=0.5$, $M =1.4 M_{\odot}$, $\alpha = 0.1$, and $r_{\rm
in}=7$.  The numerical values shown are for $\xi = 1$ and should be
multiplied by $\xi$ for other values of this parameter. Because we
have taken $\bar{t}=1$ s, the vertical range in the plot corresponds
to a total fallback mass ranging from $0.1M_\odot$ at the bottom to
$1M_\odot$ at the top.  Even for this fairly modest mass budget, we
see that the outflow energy is quite significant.

Next, we attempt a more realistic calculation using numerical
simulations of core-collapse supernovae as a guide.  Livne (2004) has
presented results corresponding to the collapse of a $11M_\odot$
progenitor star.  At a particular time, the shock is at $R = 240$ km
($r_{0}$ = 57 for a compact core of mass $M$ = 1.4 $M_{\odot}$), and
the fluid velocity, density, pressure and temperature on the two sides
of the shock are as in Table~\ref{table:shock}. Let us call the
upstream (downstream) region of the shock as ``region I'' (``region
II''). Hereafter, the subscript I (II) refers to physical quantities
in the particular region.

\begin{deluxetable}{cllll}
\tablecolumns{5}
\tablewidth{0pc}
\tablecaption{
Physical quantities around the
shock front at $R = 240$ km in the numerical simulations of
core-collapsed supernova. In this case, the progenitor star has an eleven
solar mass (Liven 2004).
}
\tablehead{
\colhead{} & \colhead{$v~(\cm~\s^{-1})$}
& \colhead{$\rho~(\g~\cm^{-3})$}
& \colhead{$p$~($\erg~\cm^{-3}$)}
& \colhead{$T$~(K)}
}
\startdata
 upstream (I-region) &
-3.0 $\times10^{9}$ &
5.0 $\times 10^{7}$ &
2.0 $\times 10^{25}$ &
6.4 $\times10^{9}$
\\
 downstream (II-region) &
-2.0 $\times 10^{8}$&
3.5 $\times 10^{8}$&
8.0 $\times 10^{26}$&
1.3 $\times 10^{10}$
\\
\enddata
\label{table:shock}
\end{deluxetable}

Using the Rankine-Hugoniot relation for the product of the velocity
and the matter density, which is defined in the rest frame of the
shock front, we find that
\begin{eqnarray}
    \label{eq:RH_relation}
    v_{\rm rest}~\rho_{\rm rest}= (v_{\rm I} - v_{\rm sh}) \rho_{\rm I} =
    (v_{\rm II} - v_{\rm sh})
    \rho_{\rm II},
\end{eqnarray}
where $v_{\rm sh}$ is the velocity of the shock front in the rest
frame of the central star. Substituting the values in
Table~\ref{table:shock} into Eq.~(\ref{eq:RH_relation}), we see that
$v_{\rm sh} = 2.7 \times 10^{8} \cm~\s^{-1}$. That is, the shock front
is not completely stalled but is gradually moving outward.  The mass
accretion rate can be roughly obtained by $\dot{M}_{0} \sim 4 \pi
r_{0}^{2} \rho_{\rm rest}~v_{\rm rest} \sim 0.6 M_{\odot} \s^{-1}$.
Using the analysis presented earlier in this subsection, we can
estimate the total time-integrated outflow energy $E_{w}$ if the shock
front is stationary. For instance, for $(r_{0}, \dot{M}_{0} )=(57, 0.6
M_{\odot} \s^{-1})$, Fig.~\ref{fig:tinteg_panel}~(a) gives $E_{w} \sim
3 \times 10^{51}$~erg, if we take $\bar{t}=1$ s.

However, since we have estimated the outward speed of the shock and
know that it is not zero, let us also calculate the results for this
particular case.  Although we do not know the exact behavior of the
shock front as a function of time, numerical simulations of
core-collapse supernovae generally show that the shock front stalls
after a short time $\lesssim 1$~s. Therefore, for simplicity, we
assume that the velocity of the shock front evolves as
\begin{eqnarray}
    \label{eq:vsh_t}
    v_{\rm sh}(t) = v_{\rm sh}(0)
    \exp\left(-t/\bar{t} \right),
    \qquad {\rm for}~t \ge 0,
\end{eqnarray}
where we use the same $\bar{t}$ as in Eq.~(\ref{eq:M_0_t}). Thus, the
position of the shock front evolves as
\begin{eqnarray}
    \label{eq:r_0_evolv}
    r_{0}(t) &=& r_{0}(0) + \int_{0}^{t} v_{\rm sh}(t) dt \nonumber \\
    &=& r_{0}(0) + v_{\rm sh}(0)~\bar{t}~\left[1 - \exp\left(-t/\bar{t}
    \right) \right],
\end{eqnarray}
where $r_{0}(0)=57$ and $v_{\rm sh}(0) = 2.7\times10^{8} \cm~\s^{-1}$.

In Fig.~\ref{fig:tinteg_panel}~(b), we plot the time-integrated total
outflow energy $E_{w}$ as a function of $s_{0}$ for three choices of
$\bar{t}$: 0.5, 1, and 2 s. From the figure, we see that the outflow
energy is $E_{w} \sim 10^{51-52}\xi$~erg over a wide range of values
of $s_{0}$.~\footnote{Because we terminated the numerical calculation
for $R > 10^{8} \cm$, these numerical values are conservative lower
bounds on the outflow energies. However, we checked that, even if we
include the contribution from $R > 10^{8}\cm$, it does not change the
results significantly.} The total mass supplied to the fallback disk
is $0.3M_\odot$, $0.6M_\odot$, $1.2M_\odot$, respectively, for the
three choices of $\bar{t}$ considered in this calculation. This is
quite conservative because we expect much more mass to be available in
a failed supernova. Even so, and even if we make the conservative
assumption that $\xi\sim0.1$, we see that quite a lot of energy is
expected in the outflow.

One additional point is that, in the delayed supernova scenario, the
continued longtime accretion will tend to increase the mass of the
remnant neutron star.  This may help resolve a problem in some classes
of simulations (Herant et al. 1994; Burrows et al. 1995; Scheck et
al. 2004), viz., that the mass of the neutron star predicted in these
simulations is less than 1.4 $M_{\odot}$.

\subsection{Nuclear Reactions in the Outflow}
\label{subsec:nuclear}

The outflowing material in the disk outflow begins as dissociated
nucleons, but as it flows out and cools, it will undergo nuclear
reactions of various kinds (MacFadyen 2003). We briefly touch on some
interesting phenomena that may be expected as a result.

As discussed in \S 2.5 and shown in
Fig.~\ref{fig:xnuc_np_p_ne_panel}~(a), the photodissociation of nuclei
in the accretion flow acts as an effective cooling term.
Correspondingly, the recombination of nucleons into nuclei in the
outflow is exothermic and increases the energy available in the
outflow. The energy release is about 8 MeV per nucleon (MacFadyen
2003), which is substantial. This energy was not included in the
energy estimates obtained in the previous subsections.
Figure~\ref{fig:E2dotcontour_panel} shows how those results are
modified when the recombination energy is included. In the new
calculations, the outflowing gas at each radius is assumed to start
off with $X_{\rm nuc}$ equal to the value given in
equation~(\ref{eq:Xnuc}), and the energy release from this material is
calculated assuming that each free nucleon releases 8 MeV through
recombination.  The particular case shown is for $\xi=0.1$.  A
comparison with Figure~\ref{fig:Edotcontour_panel} shows that the
recombination energy is quite substantial.  Without including this
energy, for $\xi=0.1$ the maximum efficiency of the outflow for the
optimum choice of $r_0$ and $\dot m_0$ is only $\dot E/\dot M_0c^2
\sim 10^{-3}$, whereas with the recombination energy included
the efficiency can become as high as $8\times10^{-3}$.

Another interesting possibility is suggested by
Fig.~\ref{fig:xnuc_np_p_ne_panel}~(b), where we see that the neutron
to proton ratio in the disk is quite high over large regions of
parameter space.  Such a neutron rich region with a short dynamical
timescale ($\ll 1~\s$) is likely to be the site of $r$-process
nucleosynthesis (Hoffman, Woosley \& Qian 1997).~\footnote{For the
neutron capture reaction to be more rapid than $\beta$-decay, we need
a large neutron flux $(\gtrsim 10^{31} \cm^{-2} \s^{-1})$. Our model
easily satisfies this condition.}  The interesting thing in our model
is that the neutron rich material flows out as part of the
advection-induced wind. During the outflow, $r$-process elements could
be generated and these will be ejected by the supernova
explosion. This possibility has been explored by various authors
within the context of gamma-ray bursts and the collapsar model (Pruet
et al.  2003, 2004a,b; Fujimoto et al. 2004; Surman \& McLaughlin
2004, 2005).  Our proposal is that similar considerations should apply
also in the context of regular supernovae.

It should be noted that there is at present no accepted model for
successful $r$-process nucleosynthesis (see recent papers by Qian
\& Wasserburg 2002 and Qian 2005).  From the viewpoint of event
rates, the core-collapse supernova is a more likely site for
$r$-process nucleosynthesis than the binary neutron star merger
(Eichler et al. 1989; Freiburghaus et al. 1999). This is because
even in metal poor stars with Fe/Fe$_{\odot}$ = $10^{-4}$ --
$10^{-3}$, typical $r$-process elements such as Eu have been
discovered (see McWilliam et al. 1995).  This means that
$r$-process nucleosynthesis must happen on the same timescale as
Fe production in supernovae.

Standard explosive nucleosynthesis in core-collapse supernovae can
produce only the lighter $r$-process elements with mass number $A <
130$, such as Rh and Ag.  Some other mechanism is needed to produce
the heavier $r$-process elements with $A > 130$. Moreover, this
mechanism should not produce too much light $r$-process elements.  If
it did, we would have difficulty explaining the fact that, among metal
poor stars, there is a variation of the abundance of the heavier
$r$-process elements with $A > 130$ whereas the abundances of the
elements from O to Ge and the lighter $r$-process elements up to Ag
are unchanged (Qian 2005).  In standard models of core-collapsed
supernovae, neutron rich regions suitable for $r$-process synthesis do
exist, but the problem is that these regions generally do not flow out
but rather fall back on the proto-neutron star.  Although the
possibility that a neutrino driven wind may eject the neutron rich
region has been studied, at least one of the following conditions has
to been satisfied: (i) low lepton fraction $Y_{e}$, (ii) short
dynamical timescale, and (iii) large entropy per baryon. To our
knowledge, these conditions have not been achieved naturally in
supernova wind models (Hoffman, Woosley \& Qian 1997)~\footnote{Some
attempts have been made to resolve the above difficulties making use
of special assumptions, e.g., neutrino oscillations between active and
sterile neutrinos (Fetter et al. 2003 and references therein),
jet-like MHD explosion (Nishimura et al. 2005) for a lower lepton
fraction, general relativistic effect for a shorter dynamical
timescale (Otsuki et al. 2000), quantum electrodynamical effect in the
strong magnetic field for a larger entropy per baryon (Kohri, Yamada
\& Nagataki 2004), and so on.}.

Another idea that has been considered by some workers is
accretion-induced collapse (AIC) as a candidate site for $r$-process
nucleosynthesis (Qian \& Wasserburg 2002).  It is reported that this
scenario produces heavier elements with $A > 130$ without
overproducing the lighter elements.  We cannot judge whether or not
the necessary conditions are naturally realized.

In comparison to these other models, the disk wind model we propose
here appears to be plausible and quite natural.  The main selling
point of this model is that we can easily bring the neutron rich
material and $r$-process elements out during the core-collapse
supernova explosion.

\section{Summary and Discussion}

In this paper we have shown that, if an accretion disk --- an NDAF
--- forms around a proto-neutron star during core collapse of a
massive progenitor star, then a substantial quantity of mass is
likely to be ejected from the disk and to carry with it a large
amount of mechanical energy. The outflow energy could be as much
as $10^{51}$ erg in favorable situations and might be sufficient
to convert a failed supernova explosion into a successful one. A
virtue of this proposal is that essentially all the energy is
available to power the explosion. This is in contrast to
neutrino-driven supernovae where only a small fraction  of the
energy is deposited in the stellar mantle.

We have estimated the energy in the outflow corresponding to two
distinct scenarios. In one scenario (\S~\ref{subsec:application1},
Fig.~\ref{fig:tinteg_prompt_panel}), we assume that the disk forms as
part of the initial core collapse and that the outflow induces a
prompt supernova explosion. In the second scenario
(\S~\ref{subsec:application2}, Fig.~\ref{fig:tinteg_panel}), we assume
that there is no prompt explosion and that the shock stalls.  However,
some of the subsequent fallback material forms a disk whose outflowing
wind re-energizes the shock and causes a delayed supernova explosion.

Depending on the parameters we assume, primarily the amount of mass
available in the disk, both the above scenarios are feasible for
producing a supernova explosion. However, the second scenario (which
leads to a delayed explosion) appears to be more promising for two
reasons. First, the total amount of fallback mass in a failed
supernova can be several solar masses (essentially all the mass in the
envelope of the progenitor star) whereas the mass in the initial
prompt disk is likely to be no more than a few tenths of a solar mass.
Thus, the mass and energy budget is much larger in the delayed
scenario.  Second, the delayed fallback material is likely to have
larger specific angular momentum since it originates from farther out
in the progenitor star. Thus, this material is more likely to form a
disk.

The results shown in Figures~\ref{fig:tinteg_prompt_panel} and
\ref{fig:tinteg_panel} are conservative in that they do not include
the additional energy that is released when the dissociated nucleons
in the outflowing gas recombine to form larger nuclei. The latter
process releases about 8 MeV per nucleon (MacFadyen 2003) which can be
a substantial fraction of the total outflow energy for certain
parameter choices. Figure~\ref{fig:E2dotcontour_panel} shows how the
results shown in Figure~\ref{fig:Edotcontour_panel} are modified when
the recombination energy is included.  The difference is quite large,
especially when the energy efficiency factor $\xi$ of the outflow is
small, say $\sim0.1$. This makes the scenario proposed in this paper
even more promising.

Although we have taken pains to model the physics of the NDAF in as
much detail as possible, the numerical results we have obtained are
still only crude estimates. This is because there are several large
uncertainties in the model which we have had to absorb in various free
parameters.  In both the prompt and delayed scenarios, we have no firm
estimate of how much mass goes into the disk. The mass depends on the
angular momentum distribution of the progenitor star, which is poorly
understood.  Also, we do not know how much of the mass in the disk
flows out in the wind. While it is generally understood that
advection-dominated accretion flows are likely to have strong outflows
(Narayan \& Yi 1994, 1995a), the exact amount of outflow is
uncertain. We have included an index $s_0$ plus an approximate
prescription for how the index varies with the degree of
advection-domination in the accretion flow (see
eq.~(\ref{eq:S_R})). Finally, we do not have a precise estimate of the
specific energy of the outflowing gas. We make the plausible
assumption that the energy is a fraction $\xi<1$ of the escape energy,
but there are no strong constraints on the value of $\xi$.

Despite these large uncertainties, we believe our calculations are
realistic enough to demonstrate that there is likely to be
considerable energy in the outflow.  One question to ask is: Where
exactly does the energy come from?  The responsible agency is gravity,
helped by viscosity (or more accurately shear stress in the accretion
flow). As is well-known (e.g., Frank et al. 1992), viscosity has three
important effects on an accretion flow.

(i) Viscosity induces a shear stress which transports angular momentum
outward, enabling mass to flow in. This is obviously critical for the
whole accretion process to occur in the first place.  The
``viscosity'' is likely to be generated via the magnetorotational
instability (Balbus \& Hawley 1991).

(ii) Viscous dissipation heats up the gas locally. If this heat is
radiated immediately, then the gas remains cold and its binding energy
is approximately $GM/2R$ (in the Newtonian limit), i.e., it is equal
to half the local potential energy.  Such highly bound gas is not
easily ejected and we do not expect a significant wind from the
disk. However, if the heat energy is not radiated, i.e., if the flow
is advection-dominated, then the gas is much more loosely bound to the
central mass; in fact, the gas is actually unbound as we discuss below
in (iii). An ADAF is thus expected to have heavy mass loss, as
originally highlighted by Narayan \& Yi (1994, 1995a).

(iii) Finally, viscosity transports energy outward.  The outward
energy flux is given by $F_{\rm out} = T_{r\phi}\Omega$, where
$T_{r\phi}$ is the local shear stress (which transports angular
momentum outward) and $\Omega$ is the local angular velocity of the
gas. In the case of an ADAF, this outward transport of energy is
extremely important. As we discussed above in (ii), the gas in an ADAF
does not lose any of its initial binding energy because it is
radiatively inefficient. Now, we see that, in addition, it acquires
extra energy from gas further in through viscous energy transport. The
net result is that the gas in an ADAF ends up with {\it positive}
energy (or negative binding energy). Narayan \& Yi (1994, 1995a)
explicitly demonstrated this by estimating the Bernoulli parameter of
the gas and showing that this quantity is positive for a self-similar
ADAF.  The positive energy drives the outflow and deposits energy in
the surroundings.  Where does the energy come from?  Ultimately, it
comes from the gravitational potential energy released by the gas that
falls onto the compact star on the inside.

The above discussion establishes the strong connection between
advection-dominated accretion and outflowing winds.  Therefore, the
mechanism we have described will not operate unless the accretion flow
is strongly advection-dominated.  Fortunately, ADAF-like conditions
are present over a wide range of parameter space in an NDAF (see
Fig.~\ref{fig:f}).  Equally, the mechanism will not operate in a
numerical simulation of a supernova explosion unless one includes
viscosity self-consistently and keeps track of the energy that is
transported and dissipated viscously.

To the best of our knowledge, viscous interactions have not been
properly incorporated in most supernova simulations done so far.  An
exception to this statement is the work of Fryer \& Heger (2000) which
investigated rotating collapse with the inclusion of
$\alpha$-viscosity.  While there is some evidence for outflowing gas
in their simulation (see their Fig. 14), nevertheless, the energy
carried by the wind is apparently not very important because the
authors find that the supernova explosion develops more slowly in a
rotating star compared to a non-rotating star.  Without knowing all
the details of the simulation, it is difficult to say why this work
did not find the energetic wind that we predict.  Values of $\alpha$
over the range 0.1 to 0.0001 were apparently used and this may be a
clue.  Comparing Figures 8 and 9 of the present paper, we see that
when $\alpha$ is reduced from 0.1 to 0.01, the range of $\dot M$ over
which the wind is relatively inefficient (because the accretion flow
is relatively efficient at radiating neutrinos) increases
substantially.  The reason for this is explained in \S~3.2.  For yet
smaller values of $\alpha$ such as those mentioned in Fryer \& Heger
(2000), the effect would be even more enhanced.  It is thus possible
that the viscosity employed by Fryer \& Heger (2000) was too weak to
exhibit the effects we have described.  It is also possible that the
progenitors they used did not rotate fast enough to develop a
substantial accretion disk around the proto neutron star.

Regardless, once three-dimensional simulations of rotating collapse
are done with full MHD (so that the magnetorotational instability is
able to develop and provide ``viscosity'' self-consistently), the
effects we have described in this paper ought to be seen. The work of
Proga et al. (2003) and Moseenko et al. (2005) is a beginning in this
direction.

It should be noted that the outflows we consider here are distinct
from the relativistic jets that are popular in collapsar models of
gamma-ray bursts (MacFadyen \& Woosley 1999) or that are invoked in
some models of supernovae (Wheeler et al. 2002).  Our outflows are
relatively slow.  Even the gas that is ejected from the innermost
region of the disk has a speed only $\sim0.2c$ (its kinetic energy is
$\xi GM/R_{\rm in}$); the gas that comes out from larger radii is even
slower.  Also, we do not expect the outflow to be highly collimated as
visualized in jet models. In our view, if at all an ultrarelativistic
jet is present, it is likely to be produced by some mechanism that is
completely different from the disk outflow we have considered. The
mechanism will probably be related to the compact object in the middle
and not the disk.  Of course, a fraction of the jet energy may
contribute to re-energizing the stalled shock (Wheeler et al.  2002)
and may assist the disk outflow in producing the explosion.  Equally,
the disk outflow that we discuss here may help, or even
play an important role, in the collimation of the relativistic
jet. 

An important point to note is that the outflow discussed in this paper
has a clear direction associated with it, namely the rotation axis of
the system.  The impact of this non-spherical wind on the stalled
shock is likely to result in an asymmetric explosion for which there
is growing evidence from polarization observations of supernovae
(Wheeler et al., 2000; Wang et al., 2001).

Finally, we note that several regions in the disk have a large neutron
to proton ratio.  Because of the short dynamical timescale, these
neutron-rich regions are candidate sites for $r$-process
nucleosynthesis. In our model, the neutron-rich gas and the
$r$-process elements that it synthesizes are naturally transported out
by the outflow and ejected in the supernova explosion. The model may
thus be of interest for studies of the origin of $r$-process elements
in the universe (Pruet et al.  2003, 2004a,b; Fujimoto et al. 2004;
Surman \& McLaughlin 2004, 2005).

\section*{Acknowledgements}

We wish to thank E. Livne for providing unpublished results from core
collapse simulations, and the anonymous referee for helpful comments.
K.K. also thanks R. Sawyer, Shoichi Yamada and M. Yokosawa for useful
discussions. This work was supported in part by NASA grant NAG5-10780,
NSF grant AST-0307433, JSPS Research Fellowships 15-03605 and
US-Israel BSF grant.


\vfill\eject
\begin{appendix}

\section{Appendix: Reaction rates and various timescales}
\label{sec:timescale}

Here we compare the timescale of the various scattering processes with
the dynamical timescale in the system.  The dynamical timescale is
represented by the accretion time,
\begin{eqnarray}
    \label{eq:tacc}
    t_{\rm acc} &\equiv& \frac1{\alpha}\sqrt{\frac{R^3}{G
    \mbh}}\left( \frac{R}{H}\right)^2.
\end{eqnarray}
For reference, if we simply assume that the disk half-thickness is $H
\sim R/2$ (as appropriate for an ADAF), then the dynamical timescale
is approximately given by $t_{\rm acc} \simeq 3.0 \times 10^{-2} \ \s
\left({\alpha}/{0.1}\right)^{-1}\left({R}/{7 R_{\rm
S}}\right)^{3/2}\left({\mbh}/{3 M_{\odot}}\right)$.

The timescale of the scattering rates among photons and electrons is
much faster than the accretion timescale because of the
electromagnetic interaction (Thomson scattering), i.e., $t_{\gamma e}
\sim 1/(\sigma_{\rm T}n_{\gamma} c) \sim 10^{-18} T_{11}^{-3}~\s $,
with Thomson cross section $\sigma_{\rm T}$. Therefore, we obviously
expect that photons and electrons interact each other rapidly and are
immediately thermalized.

On the other hand, neutrinos with energy $\sim k_{\rm B}T$ scatter off
the background nucleons and electrons only through the weak
interaction. The reaction rates of these scattering processes are
roughly estimated by
\begin{eqnarray}
    \label{eq:gamma_nue}
\Gamma_{\nu e} \sim \Gamma_{\nu
N} \sim G_{\rm F}^{2} (k_{\rm B}T)^{2} n_{e}.
\end{eqnarray}
Note that this rate is of the same order of magnitude as that of
neutrino-antineutrino pair production by background electron-positron
annihilation $\Gamma_{e^{+}e^{-} \to \nu \anti{\nu}}$. Thus, the
timescale of scattering (or the pair-production) is $t_{\nu e} \sim
t_{\nu N} \sim t_{e^{+}e^{-} \to \nu \anti{\nu}} \sim \left[G_{\rm
F}^{2} (k_{\rm B}T)^{2} n_{e} \right]^{{-1}}$.  The condition $t_{\nu
e} \ll t_{\rm acc}$, i.e., $\Gamma_{\nu e}/t_{\rm acc}^{-1} \gg 1$
approximately means that neutrinos can scatter off background
particles and transfer their energy to them within a typical dynamical
timescale. In Fig.~\ref{fig:timesale_panel}~(a) we plot contours of
the ratio $\Gamma_{\nu e}/t_{\rm acc}^{-1}$. We find that neutrinos
scatter off background particles efficiently in the upper left
region. However, please note that in terms of the thermalization of
neutrinos, $\Gamma_{\nu e}/t_{\rm acc}^{-1} \gg 1$ is just a necessary
condition, not a sufficient condition.  In this study we
conservatively regard the $\nu_{i}$'s to be thermalized only when both
$\tau_{a,\nu_{i}} \gg 1$ and $\tau_{s,\nu_{i}} \gg 1$ are
satisfied. As shown in Figs.~\ref{fig:tau_nue_panel}~(a) and (b), the
parameter region where both conditions are satisfied is surely
included in the region where $\Gamma_{\nu e}/t_{\rm acc}^{-1} \gg 1$
in Fig.~\ref{fig:timesale_panel}~(a).

It is also useful to discuss the timescale of the interconverting
rates between neutrons and protons through the weak interaction. The
expressions of the rates are presented in
Eqs.~(\ref{eq:beta_reac1})~--~(\ref{eq:beta_reac6}). In particular,
electron capture by a proton and positron capture by a neutron are
important to interconvert neutrons and protons in the electromagnetic
thermal bath. This is because these processes are effective even when
the neutrino-nucleon scattering in
Eqs.~(\ref{eq:beta_reac4})~--~(\ref{eq:beta_reac6}) is ineffective due
to insufficient thermalization of background neutrinos. We plot
contours of the ratio $\Gamma_{pe^{-}\to n\nu_{e}}/t_{\rm acc}^{-1}$
in Fig.~\ref{fig:timesale_panel}~(b). From the condition
$\Gamma_{pe^{-}\to n\nu_{e}}/t_{\rm acc}^{-1} \gg 1$ (which is the
same as $(n/p)\times \Gamma_{ne^{+}\to p\anti{\nu}_{e}}/t_{\rm
acc}^{-1} \gg 1$ in thermal equilibrium of $n/p$), we see that the
electron-capture processes by nucleons are effective in the upper left
region of the $r$-$\dot{m}$ plane.

\section{Appendix: Actual calculation of $n/p$}
\label{sec:actural_np}

Here we discuss the approximations we use to calculate the neutron to
proton ratio $n/p$ in this study. Ideally, we should solve the balance
equation of the interconverting reactions between neutrons and protons
written down in Eqs.~(\ref{eq:dnpdt}) and (\ref{eq:dnndt}). However,
let us make some bold but reasonable approximations since the exact
calculations require intensive computations and are not needed for our
current purpose.

We classify the $r$-$\dot{m}$ plane into following three regions: (i)
the region where neutrinos are completely thermalized and the
timescale for the interconverting reaction $\Gamma_{n\leftrightarrow
p}$ is much smaller than the dynamical timescale, (ii) the region
where $\Gamma_{n\leftrightarrow p}$ is rapid but the the
thermalization of neutrinos is incomplete and the neutrino-nucleon
scattering is unimportant, and (iii) the region where
$\Gamma_{n\leftrightarrow p}$ is slow compared to the dynamical
timescale. We discuss each of the three cases in detail.

Fist of all, it is necessary to know the distribution functions of
electrons and neutrinos, which govern the balance equations of the
interconverting reactions.  As we have discussed in
Appendix~\ref{sec:timescale}, the thermalization of electrons is
easily realized because the electromagnetic interaction is very rapid
compared to the dynamical timescale.

\begin{figure}
\epsscale{1.0}
\plotone{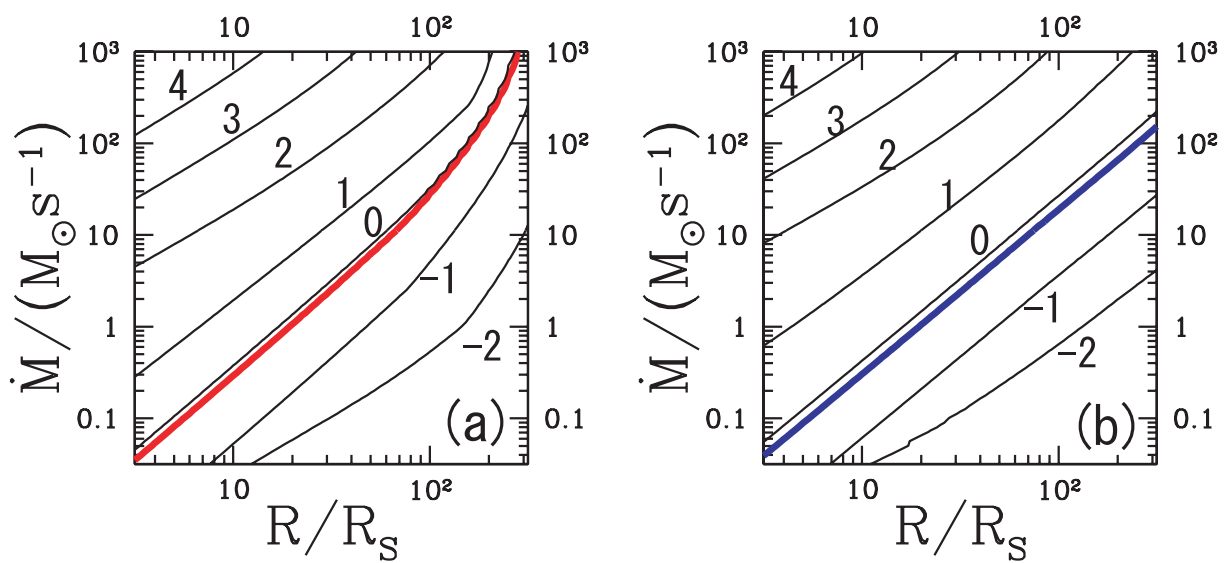}
\vspace{.0in}
\caption{ (a) Contours of the neutrino absorption optical depth
$\log_{10}(\tau_{a,\nu_{e}})$ in the $r$--$\dot{m}$ plane. The thick
solid line corresponds to $\tau_{a,\nu_{e}} = 2/3$. (b) Contours of
the neutrino scattering optical depth $\log_{10}(\tau_{s,\nu_{e}})$.
The results are for $\alpha = 0.1$ and $M=1.4 M_{\odot}$.  }
\label{fig:tau_nue_panel}
\end{figure}

On the other hand the thermalization of neutrinos is not obvious. The
problem is that neutrinos scatter off the background particles such as
electrons, nucleons or neutrinos only through the weak
interaction. For accurate results, we would have to solve a set of the
Boltzmann equations to trace the time evolution of the
energy-distribution function of neutrinos and their energy
transfer. Instead, for simplicity, we adopt the following approximate
method which does not involve explicitly solving the Boltzmann
equations. When both $\tau_{a,\nu_{i}}$ and $\tau_{s,\nu_{i}}$ are
much larger than $2/3$, it is reasonable to expect that neutrinos are
completely thermalized, so we assume this. The corresponding parameter
regime is clearly shown in Figs.~\ref{fig:tau_nue_panel}(a)~and~(b)
for the electron neutrinos. In addition, from
Fig.~\ref{fig:timesale_panel}~(a), we see that the timescale of such
scatterings through $\nu e$ ($\nu N$, and $\nu\nu$) is much more rapid
than the dynamical timescale.  Therefore, as shown in
Eq.~(\ref{eq:npratio_EQ}), we obtain $n/p = \exp(-Q/k_{\rm
B}T+\eta_{e})$ as the thermal equilibrium value of the neutron to
proton ratio (case~(i)).

If the electron neutrinos are not thermalized, i.e., if the optical
depths $\tau_{a,\nu_{e}}$ and $\tau_{s,\nu_{e}}$ are less than 2/3 in
the current context, it would be reasonable to regard that the
neutrino-nucleon scatterings in
Eqs.~(\ref{eq:beta_reac4}),~(\ref{eq:beta_reac5})~and~(\ref{eq:beta_reac6}),
are not important and can be ignored, compared with the other
processes in
Eqs.~(\ref{eq:beta_reac1}),~(\ref{eq:beta_reac2})~and~(\ref{eq:beta_reac3}).
In Fig.~\ref{fig:timesale_panel}~(b) we find the parameter regions
where the condition $\Gamma_{pe^{-}\to n\nu_{e}} / t_{\rm acc}^{-1}
\gg 1$, i.e., $(n/p) \Gamma_{ne^{+} \to n\nu_{e}} / t_{\rm acc}^{-1}
\gg 1$ is realized, while $\tau_{a,\nu_{e}}$ and $\tau_{s,\nu_{e}}$
are less than 2/3. In this region, we approximately have $n/p$ $\simeq
(\Gamma_{pe^- \to n \nu_e})/ (\Gamma_{ne^+ \to p \anti{\nu}_e} +
\Gamma_{ n \to p e^- \anti{\nu}_e})$ as the equilibrium value with
$f_{\nu_{e}} \sim f_{\anti{\nu}_{e}} \sim 0$ (case~(ii)).

Finally, if the timescale of the above interconverting reactions is
not shorter than the accretion time ,$\Gamma_{p \leftrightarrow
n}/t^{-1}_{\rm acc} < 1$, we can no longer expect any kind of
equilibrium to be achieved in the neutron to proton ratio through the
weak interaction. In this case, we assume that the neutron to proton
ratio approximately becomes unity ($n/p \simeq 1$). This is a
reasonable assumption because most of the free nucleons are produced
by the destruction (e.g., through the photodissociation) of heavy
nuclei such as He, C, N and O, which contain approximately equal
numbers of neutrons and protons (case~(iii)).

In Fig.~\ref{fig:np_T}, we show the neutron to proton ratio as a
function of the temperature in units of $10^{11}~\K$ for
representative examples of cases (i) and (ii). The dotted lines denote
case (i) in which electron neutrinos are completely thermalized and
the neutrino-nucleon collisions are important, i.e.,
$n/p=\exp({-Q/k_BT+\eta_e})$. The solid lines denote case (ii) in
which we omit the neutrino-nucleon collisions and
$n/p=\Gamma_{pe^-\rightarrow n\nu_e}/(\Gamma_{ne^+ \rightarrow p
\anti{\nu}_e}+\Gamma_{n\rightarrow pe^-\anti{\nu}_e})$. The upper
(lower) lines correspond to $\eta_{e}=1$ ($\eta_{e}=10^{-3}$).

\begin{figure}
\epsscale{1.0}
\plotone{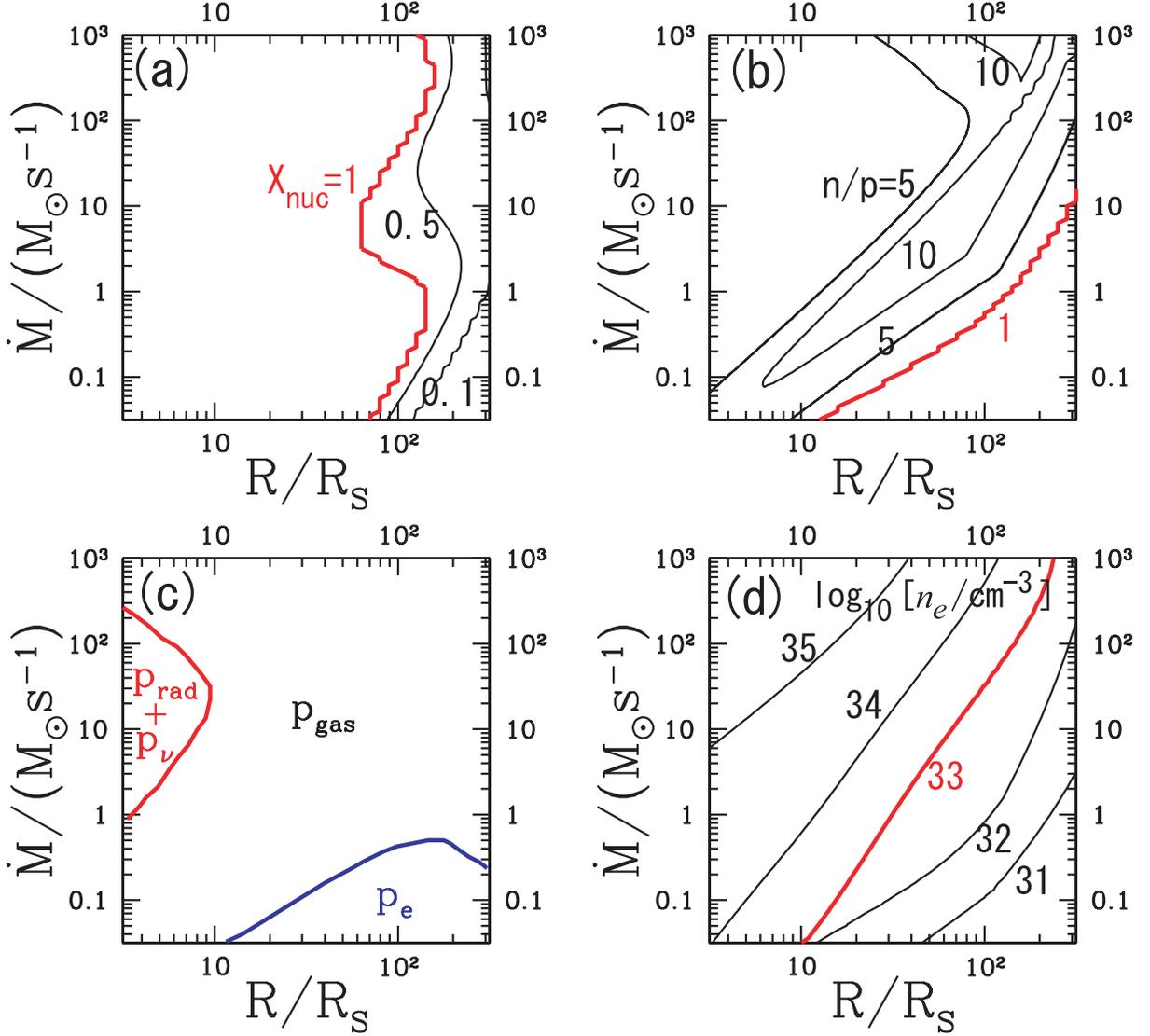}
\vspace{.0in}
\caption{ (a) Contours of the nucleon faction $X_{\rm nuc}$ in the
$r$--$\dot{m}$ plane. For $r \lesssim 150$, the region of most
interest to us, we have $X_{\rm nuc} \sim 1$, i.e., the nuclei are
completely destroyed into free nucleons.  (b) Contours of the neutron
to proton ratio $n/p$.  (c) The dominant source of pressure in
different regions of the $r$--$\dot{m}$ plane. (d) Contours of the
number density of electrons $\log_{10}[n_e/\cm^{-3}]$.  All results are
for $\alpha = 0.1$ and $M=1.4 M_{\odot}$.}
\label{fig:xnuc_np_p_ne_panel}
\end{figure}

\begin{figure}
\epsscale{1.0}
\plotone{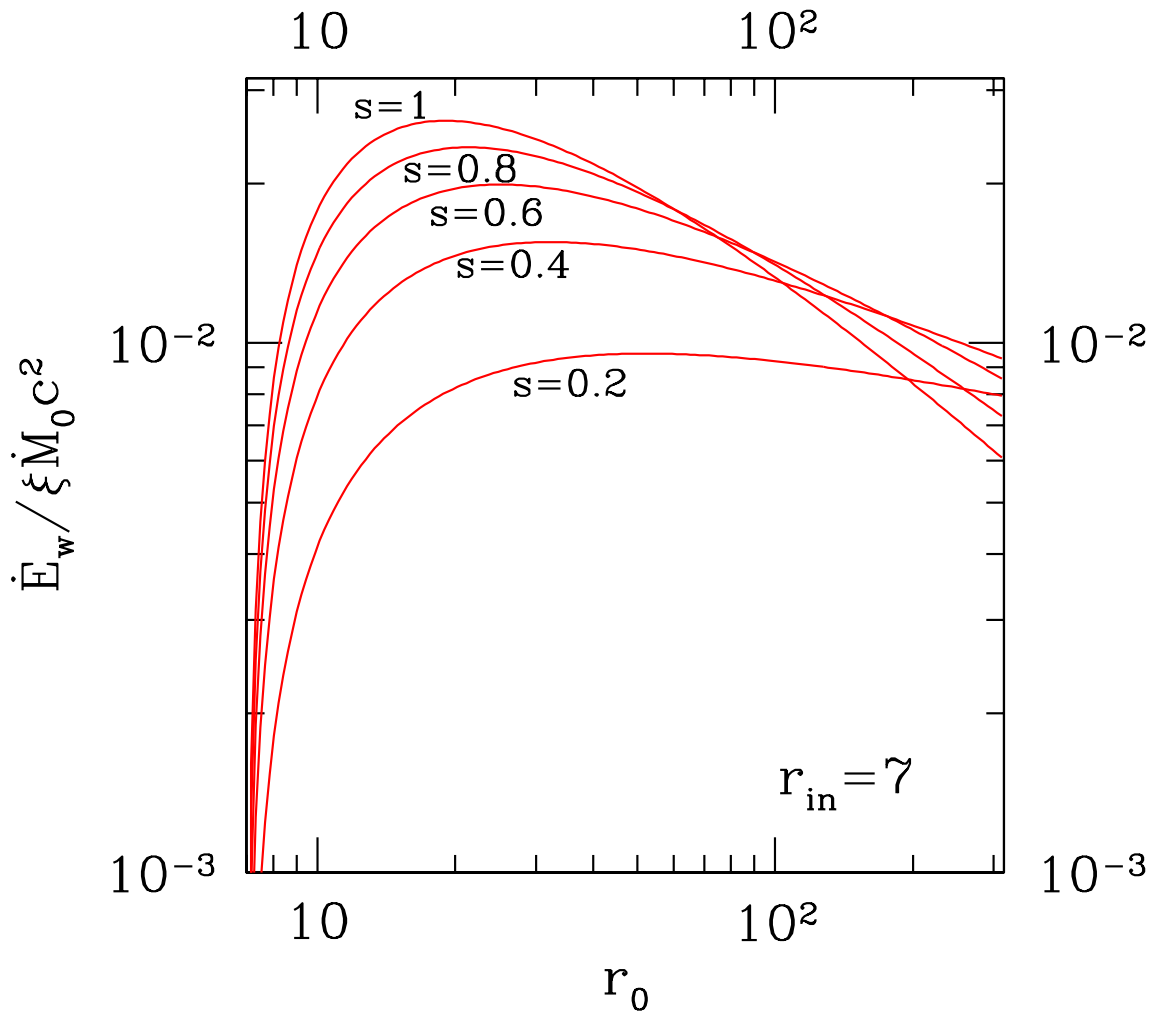}
\vspace{.0in}
\caption{Approximate analytical estimate of the energy available in
the wind from a fully advection-dominated accretion flow.  $\dot E_w$
is the power in the wind and $\dot M_0c^2$ is the rate at which rest
mass energy is supplied to the disk at its outer radius $R_0$. The
ratio of these quantities measures the outflow efficiency and is
plotted along the ordinate. The index $s$ describes how much mass is
lost in the outflow and $\xi$ measures the specific energy of the
outflowing gas.  The inner edge of the accretion flow is taken to be
at $r_{\rm in}=7$, which corresponds to $\sim 30$ km for a
$1.4M_\odot$ compact object.  (See \S~3.1 for details.)}
\label{fig:Edot0}
\end{figure}

\begin{figure}
\epsscale{1.0}
\plotone{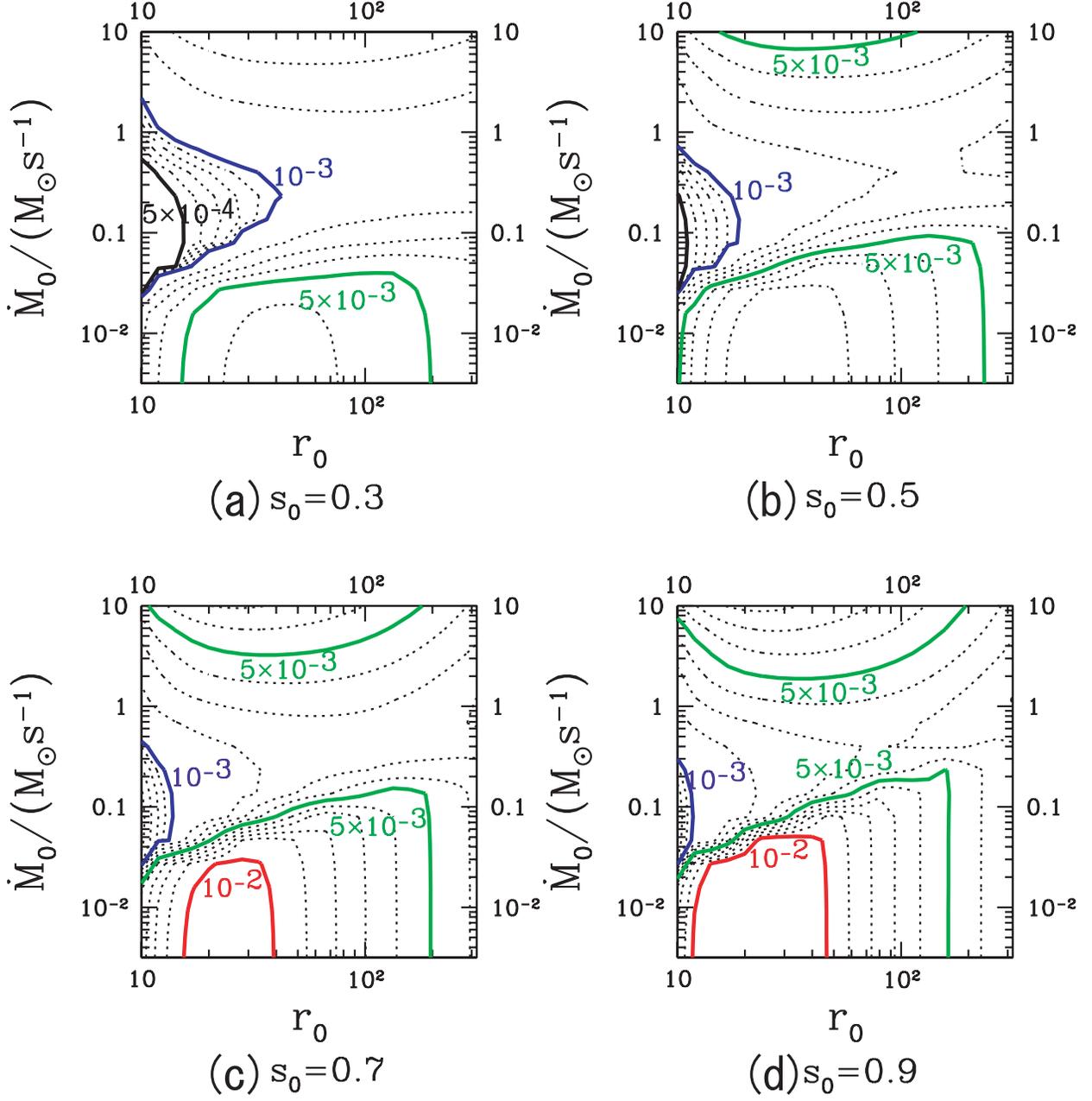}
\vspace{.0in}
\caption{ (a) Contours of the outflow efficiency $\dot{E_{w}}/(\xi
\dot{M}_{0}c^{2})$ in the $r_0$--$\dot m_0$ plane.  Here, the outflow
efficiency has been calculated more accurately than in
Fig.~\ref{fig:Edot0}, including the effect of variable advection and
using equation~(\ref{eq:S_R}) for the outflow index $s(R)$ with
$s_0=0.3$.  (b) $s_{0}=0.5$. (c) $s_{0}=0.7$. (d) $s_{0}=0.9$.  All
results are for $\alpha = 0.1$, $M=1.4M_{\odot}$ and $r_{\rm in} =
7$. }
\label{fig:Edotcontour_panel}
\end{figure}

\begin{figure}
\epsscale{1.0}
\plotone{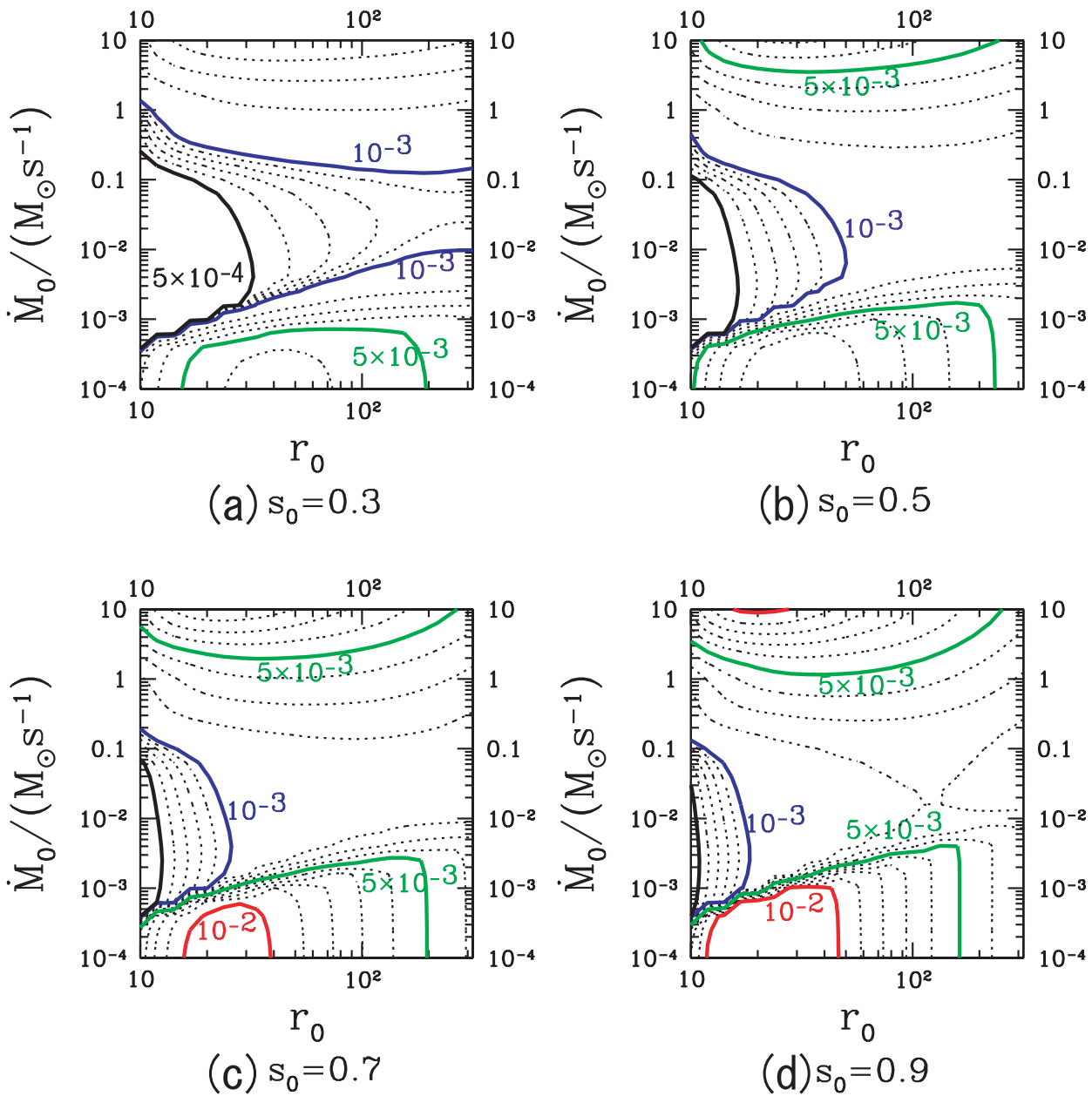}
\vspace{.0in}
\caption{ Similar to Fig.~\ref{fig:Edotcontour_panel}, but for
$\alpha=0.01$.  Note that the vertical axis extends over a wider range
of $\dot M_0$.}
\label{fig:Edotcontour_panel_a0.01}
\end{figure}

\begin{figure}
\epsscale{1.0}
\plotone{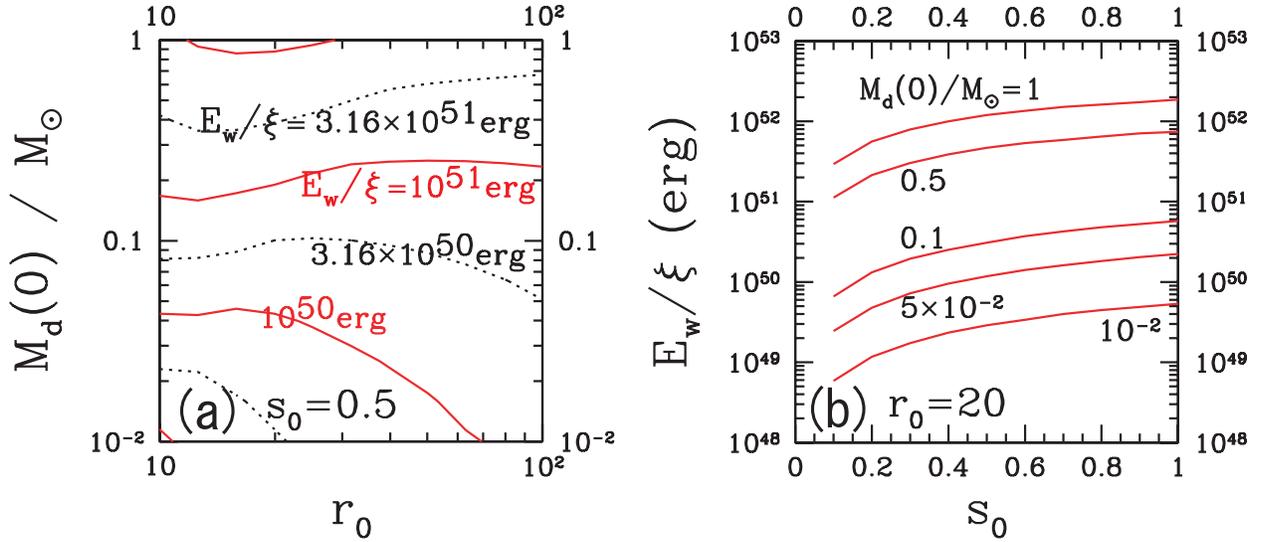}
\vspace{.0in}
\caption{ (a) Contours of the time-integrated total outflow energy
$E_{w}/\xi$ in the $r_{0}-M_{d}(0)$ plane (where $M_d(0)$ is the
initial mass in the disk) for the prompt supernova explosion scenario
(\S~3.3).  $s_{0}$ has been fixed at 0.5.  (b) $E_{w}/\xi$ as a
function of $s_{0}$ for $r_0=20$ for five choices of the initial disk
mass: $M_{d}(0)/M_{\odot}$ = $10^{-2}$, $5 \times 10^{-2}$, $0.1$,
0.5, and 1.  All results correspond to $\alpha = 0.1$,
$M=1.4M_{\odot}$ and $r_{\rm in} = 7$.  }
\label{fig:tinteg_prompt_panel}
\end{figure}

\begin{figure}
\epsscale{1.0}
\plotone{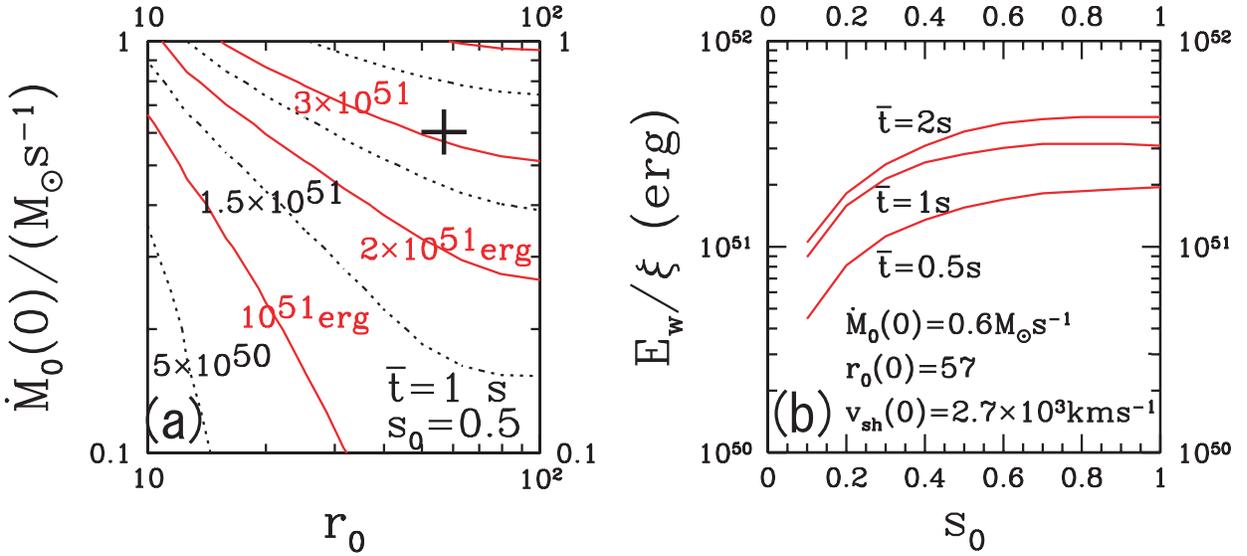}
\vspace{.0in}
\caption{ (a) Contours of the time-integrated total outflow energy
$E_{w}/\xi$ in the $r_0$--$\dot m_0$ plane for the delayed supernova
explosion scenario (\S~3.4).  The outer edge of the disk is at $r_0$
which is assumed to be fixed with time. The mass accretion rate is
initially $\dot m_0$ and decays exponentially with time with a time
constant $\bar t=1$ s.  The large cross corresponds to
$(r_{0},~\dot{M}_{0})=(57, ~0.6 M_{\odot}\s^{-1})$, a representative
example (see \S~\ref{subsec:application2}).  (b) $E_{w}/\xi$ as a
function of $s_{0}$ for three choices of the decay time constant:
$\bar t = 2, ~1, ~0.5$ s. Here, $R_0$ is assumed to increase with time
with a velocity $v_{\rm sh}(t)=v_{\rm sh}(0)\exp(-t/\bar{t})$, where
$v_{\rm sh}(0) =2.7 \times 10^{8} \cm~\s^{-1}$.  All results
correspond to $\alpha = 0.1$, $M=1.4M_{\odot}$ and $r_{\rm in} = 7$.
}
\label{fig:tinteg_panel}
\end{figure}

\begin{figure}
\epsscale{1.0}
\plotone{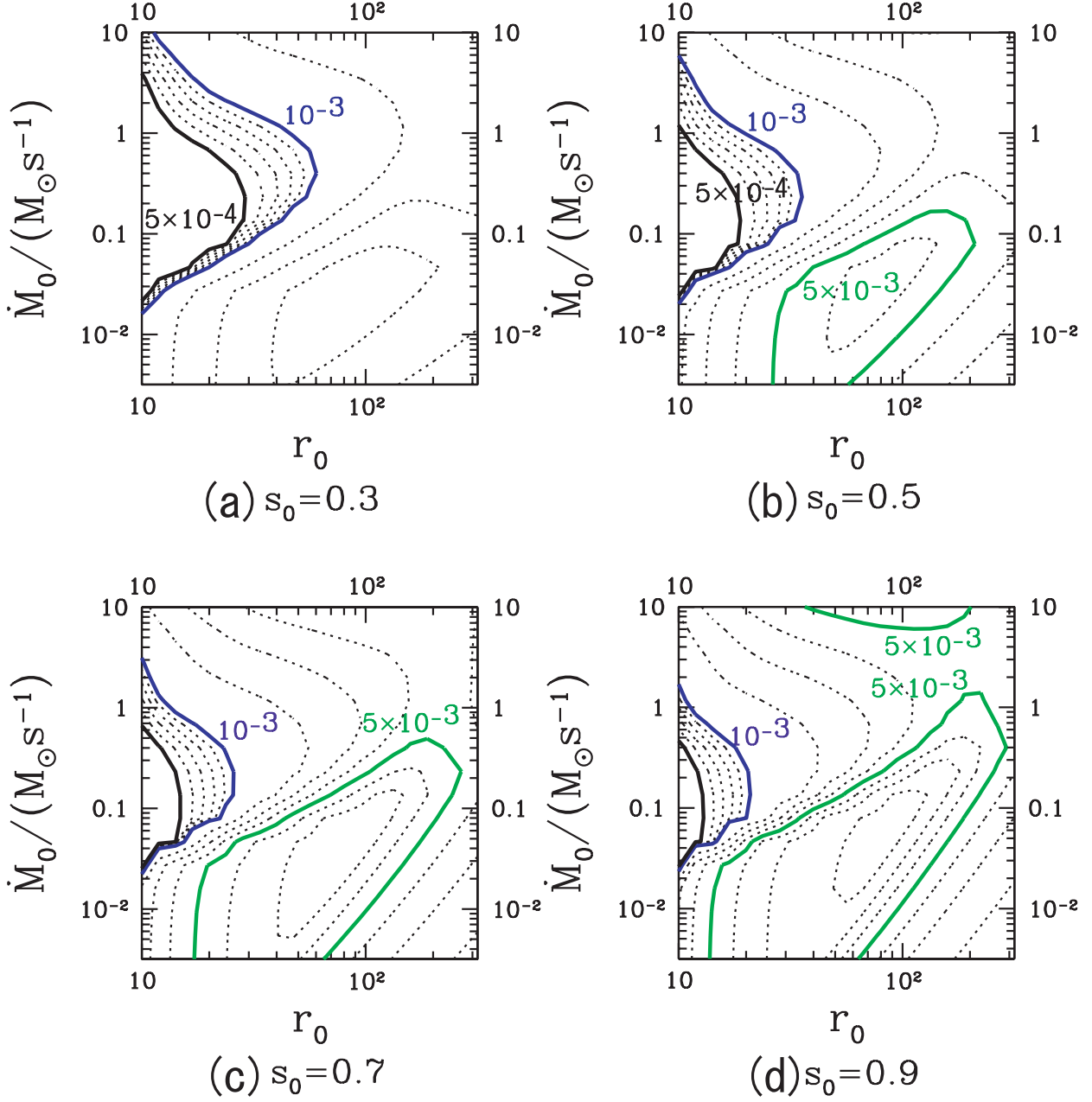}
\vspace{.0in}
\caption{ (a) Contours of the outflow efficiency $\dot{E_{w}}/(\xi
\dot{M}_{0}c^{2})$ in the $r_0$--$\dot m_0$ plane, including the
energy released through recombination of nucleons into nuclei, for
$\xi=0.1$.  Compare with Fig.~\ref{fig:Edotcontour_panel} where
recombination energy was not included.  Contour values in
Fig.~\ref{fig:Edotcontour_panel} should be multiplied by $0.1$ for the
present value of $\xi$.  For this $\xi$, the recombination energy is
seen to be highly significant.}
\label{fig:E2dotcontour_panel}
\end{figure}

\begin{figure}
\epsscale{1.0}
\plotone{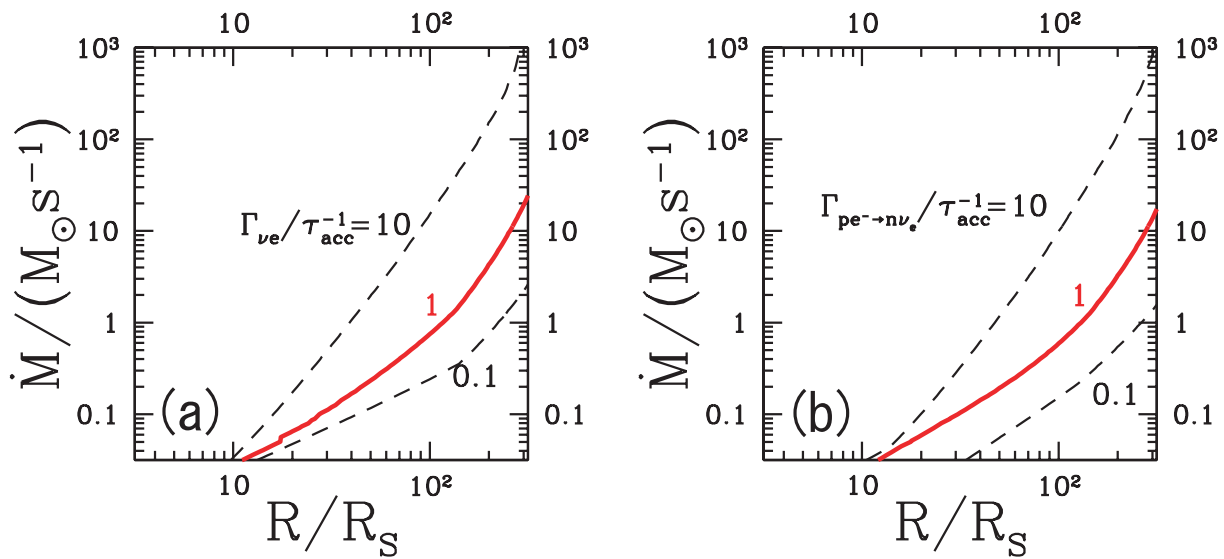}
\vspace{.0in}
\caption{ (a) Contours of the ratio of the neutrino-electron
scattering rate and the accretion rate, $\Gamma_{\nu e}/\tau_{\rm
acc}^{-1}$, in the $r$--$\dot{m}$ plane. (b) Contours of the ratio of
the proton to neutron interconversion rate to the accretion rate,
$\Gamma_{pe^{-}\to n\nu_{e}}/\tau_{\rm acc}^{-1}$.  All results
correspond to $\alpha = 0.1$ and $M=1.4M_{\odot}$.  }
\label{fig:timesale_panel}
\end{figure}

\begin{figure}
\epsscale{1.0}
\plotone{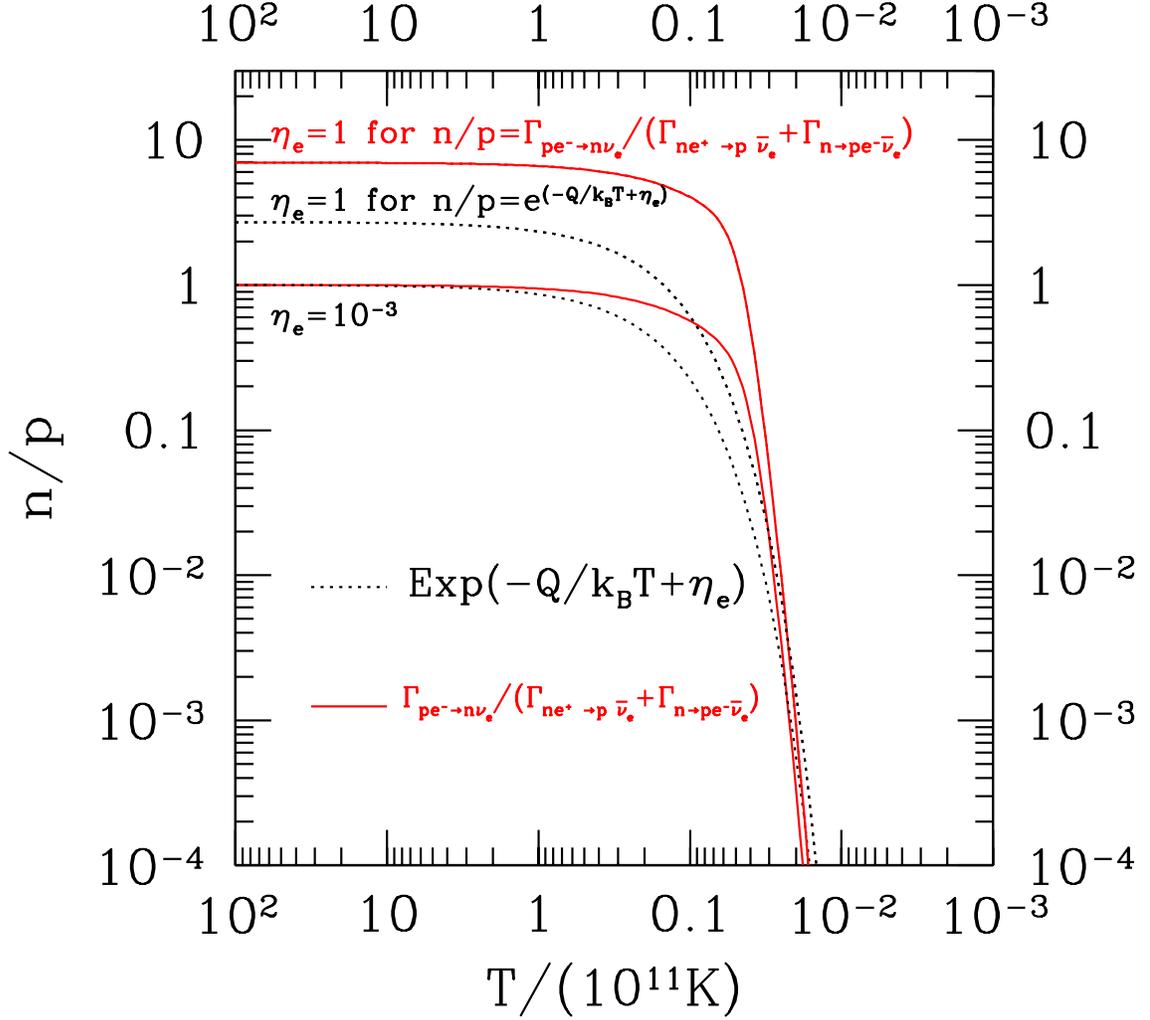}
\vspace{.0in}
\caption{ Plot of the neutron to proton ratio as a function of the
temperature.  The dotted lines represent cases in which the neutrinos
are completely thermalized and the neutrino-nucleon collisions are
important, so that $n/p=\exp({-Q/k_BT+\eta_e})$.  The solid lines
represent cases in which we omit the neutrino-nucleon collisions, and
set $n/p=\Gamma_{pe^-\rightarrow n\nu_e}/(\Gamma_{ne^+ \rightarrow p
\anti{\nu}_e}+\Gamma_{n\rightarrow pe^-\anti{\nu}_e})$). The upper
(lower) lines correspond to $\eta_{e}=1$ ($\eta_{e}=10^{-3}$).  }
\label{fig:np_T}
\end{figure}

In Fig.~\ref{fig:xnuc_np_p_ne_panel}~(b), we plot contours of the
neutron to proton ratio $n/p$ in the $r$--$\dot{m}$ plane.  We see
that $n/p$ becomes greater than unity in the upper left region. This
is typical for the equilibrium value of $n/p$ for a positive finite
$\eta_{e}$. Neutron rich material flowing out of this region could
experience $r$-process nucleosynthesis as we discuss in \S~3.5 and 4.

\end{appendix}

\clearpage

{}

\newpage

\clearpage

\clearpage

\clearpage

\clearpage

\clearpage

\clearpage

\clearpage

\clearpage

\clearpage

\clearpage

\clearpage

\clearpage

\end{document}